\documentstyle[floats,aps,epsf,amsfonts]{revtex}
\tighten
\begin{document}
\draft


\title{Gravitational self force by mode sum regularization}
\author{Leor Barack}
\address{
Department of Physics, Technion---Israel Institute of Technology,
Haifa, 32000, Israel, and\\
Albert-Einstein-Institut, Max-Planck-Institut f{\"u}r Gravitationsphysik,
Am M\"uhlenberg 1, D-14476 Golm, Germany
}
\date{\today}
\maketitle


\begin{abstract}

We propose a practical scheme for calculating the local gravitational
self-force experienced by a test mass particle moving in a black
hole spacetime. The method---equally effective for either weak or
strong field orbits---employs the {\em mode-sum regularization scheme}
previously developed for a scalar toy model.
The starting point for the calculation, in this approach, is the formal
expression for the regularized self-force derived by Mino {\em et
al}.\ (and, independently, by Quinn and Wald), which involves a worldline
integral over the tail part of the retarded Green's function.
This force is decomposed into multipole (tensor harmonic) modes, whose
sum is subjected to a carefully designed
regularization procedure. This procedure involves an analytic derivation
of certain ``regularization parameters'' by means
of a local analysis of the Green's function.
This manuscript contains the following main parts:
(1) Introduction of the mode sum scheme as applied to the
gravitational case.
(2) Two simple cases studied:
the test case of a static particle in flat spacetime, and the case of a
particle at a turning point of a radial geodesic in Schwarzschild spacetime.
In both cases we derive all necessary regularization parameters.
(3) An Analytic foundation is set for applying the scheme in more general
cases. (In this paper, the mode sum scheme is formulated within
the harmonic gauge. The implementation of the scheme in other
gauges shall be discussed in a separate, forthcoming paper.)

\end{abstract}


\section{introduction}\label{secI}

A pointlike object of mass $m$ (hereafter a ``particle'') is known to
move along a geodesic of the background spacetime in the limit where
$m\to 0$. If the mass of the particle is finite, the motion would no
longer be geodesic: interaction of the particle with its own gravitational
field will then give rise to the exertion of a ``self force'', pushing the
particle away from geodesic motion. (The non-conservative part of this
force is customarily referred to as ``radiation reaction''.)
The problem of calculating the gravitational self force is a
longstanding one. This problem is usually tackled in the context of
perturbation theory, by treating $m$ as a small parameter and looking
for the $O(m)$ correction to the geodesic equation of motion on a fixed
background.
The prototype problem---calculating the electromagnetic self force on an
accelerating (classical) electron in flat space---was considered by Dirac
in his famous 1938 paper \cite{Dirac}. Already in this flat space problem
one encounters the fundamental issue of {\em regularization}, namely, how
to correctly handle the divergence of the electromagnetic field (and self
force) at the particle's location. Dirac's
regularization yielded what is now referred to as the ``Abraham-Lorentz-Dirac''
(ALD) force (proportional to the time derivative of the electron's
acceleration). The interpretation of the ALD equation of motion and its
solutions has been subject to much further study \cite{FW}.
A formal framework for calculation of the electromagnetic self force
in {\em curved} spacetime was first developed by DeWitt and Brehme in 1960
\cite{DB}. Here, in addition to re-addressing the question of regularization,
one must also deal with the {\em non-local} nature of the self force effect:
In curved spacetime, waves emitted by the particle at a given instant may
backscatter off spacetime curvature and interact back with the particle
later on its motion. Consequently, the momentary self force acting on the
particle appears to depend on the particle's entire past history.

The continuous effort for theoretical characterization of gravitational
waveforms from strongly gravitating astrophysical systems led, in recent
years, to a rise of interest in the problem of calculating self forces in
curved spacetime. The most prominent consequence of this
has been the first development of formal tools for calculating the {\em
gravitational} self force (see below). Knowing the self force is necessary for
describing the strong field orbital evolution in various astrophysical
scenarios, in particular---the capture of a small compact object by a
supermassive black hole [an event expected to serve as a main source of
gravitational waves for the future space-based Laser Interferometer Space
Antenna (LISA)\cite{LISA}]. In some occasions, the (time-averaged) orbital
evolution, and the consequent gravitational waveforms, may be determined
by calculating energy-momentum fluxes at infinity (and through the event
horizon) and using
balance considerations\cite{balance}. This method can be applied successfully
in models where the central massive object is spherically symmetric (e.g.,
a Schwarzschild black hole), or even in the more realistic case of a
Kerr background---but then, only for equatorial orbits. In more general
cases (a non-equatorial orbit around a Kerr black hole), a full specification
of the orbital evolution requires knowledge of the temporal rate of change
of the Carter ``constant of motion'', which, to the best of our knowledge,
cannot be achieved by standard balance considerations.
In addition, balance considerations involve averaging over
orbital periods, and can only account for situations where the orbital
evolution is adiabatic. This method is inadequate in other, non-adiabatic
scenarios, as the final plunge of the particle into a black hole.
In general, therefore, it seems that one cannot avoid tackling the
problem of calculating the local, momentary self force.

In 1997, two groups---Mino Sasaki and Tanaka (MST) \cite{MST} and
Quinn and Wald (QW) \cite{QW}---have worked out (independently) a
formal framework for local gravitational self force calculations
in curved spacetime.
Their combined work presents three different derivations of the self
force, all yielding the same formal expression (though in a different
level of rigorousness). In the first approach,
MST \cite{MST} directly generalized the above mentioned analysis by DeWitt
and Brehme to the gravitational case: First, the gravitational perturbation
induced by the mass particle was evaluated near the worldline using
Hadamard's expansion of the retarded Green's function \cite{Hadamard}.
Then, the equation of motion was deduced by imposing local energy-momentum
conservation on a thin worldtube surrounding the particle's worldline.
The second approach, still by MST \cite{MST}, employed the technique
of {\em matched asymptotic expansion}, which is based on considering two
asymptotic zones outside the particle: an ``internal'' zone, where
the geometry is taken to be that of a Schwarzschild black hole with a
tidal perturbation associated with the background curvature, and an
``external'' zone, where geometry is that of a perturbation on
a fixed background. The equation of motion is then obtained by requiring
consistency of the metric in the matching region of the two zones.
The third approach for deriving the gravitational self force was presented
by QW \cite{QW},
and is based on what they called {\em the comparison axiom}. According
to this axiom, the correct self force can be deduced by appropriately
comparing the perturbation on the given curved spacetime with that on
flat spacetime. This procedure results in the elimination of the
divergent piece of the force, and, presumably, in extraction of its
correct finite part. (QW also applied their approach for calculating
the electromagnetic \cite{QW} and scalar \cite{Quinn} self forces.)

As already mentioned, all three derivations of the gravitational self
force yield the same result, which we now briefly describe.
In curved spacetime, the two-point retarded Green's function
associated with a certain wave equation (i.e., the gravitational
perturbation equation in a given gauge) is composed of two parts:
an ``instantaneous'' part, which describes the propagation of influence
along the light cone; and the so-called ``tail'' part, describing the
nonlocal effect of waves propagating {\em inside} the light cone.
The gravitational self force is derived from the gravitational
perturbation produced by the particle, which, it turn, can be expressed
as a worldline integral over the Green's function. MST and QW (MSTQW)
found that the gravitational self force (in vacuum, and with no external
forces) is due only to the {\em tail part}
of the Green's function associated with the perturbation equation in
the harmonic gauge. That is, the ``instantaneous'' part of this Green's
function yields no contribution to the self force.
[The formal expression for the gravitational self force as derived by
MSTQW is given in Eq.\ (\ref{eqII40}) below.]

We may point out here two aspects in which the gravitational self force
differs from its electromagnetic (and scalar) counterparts.
First, the gravitational self force contains no local term analogous
to the electromagnetic (or scalar) ALD force (which is proportional
to the time derivative of the particle's acceleration), as there is no
acceleration associated with the purely gravitational field.
A more fundamental difference concerns the very nature of the
gravitational self-force as a gauge-dependent entity:
When the metric is subjected to a gauge transformation (i.e.,
an infinitesimal coordinate transformation), additional terms emerge
in the particle's equation of motion, which correspond to a change in
the effective self force. (If fact, in the perturbative context the
self force is a ``pure gauge'' entity, in the sense that it can be
locally eliminated by a suitable choice of the gauge.) To provide any
meaningful gauge-invariant physical information, the self force must
therefore be supplemented by the metric perturbation to which it corresponds
(which, of course, contains all information about the gauge). In this regard,
an essential point is that MSTQW's prescription is formulated within
the framework of the harmonic gauge. A priori, it is only in this
gauge---the harmonic gauge---that the force is guaranteed to be well
defined and finite.
The analysis presented in the current paper remains within the framework
of the harmonic gauge (namely, we shall discuss the ``harmonic gauge
self-force''). A systematic study of how the self force behaves under a
change of gauge shall be presented in a forthcoming paper \cite{gauge}.

The formal expression obtained by MSTQW for the gravitational self
force was sometimes considered impractical for actual calculations,
as it was unclear how one should evaluate the nonlocal tail term
in general cases. Also, to apply this expression, one encounters
the problem of calculating the metric perturbation (and Green's
function) in the harmonic gauge, for which perturbation formalism
has not been fully developed as for other customarily used gauges
(e.g., the Regge-Wheeler or radiative gauges).
The first actual calculation of the gravitational self force based
entirely on MSTQW was recently
presented by Pfenning and Poisson \cite{EricNew}, who
considered the motion of a mass particle (as well as scalar
and electrically charged particles) moving in a weakly curved
region of spacetime. In our present manuscript, the attempt is
made to present a practical method for direct implementation of
the MSTQW result {\em in strong gravitational field}.

It should be commented that other approaches to the gravitational
self force, not
directly relying on the MSTQW analyses, were also taken recently by
a number of authors. Lousto \cite{Lousto-Letter} proposed a scheme based
on the zeta-function regularization technique, to allow self force calculation
in strong field (Lousto's scheme is similar to the method presented in this
paper, in that both methods employ a multipole mode decomposition
of the gravitational perturbation). Another method for extracting the
finite part of the self force was proposed by Detweiler \cite{Detweiler}.
Most recently, Nakano and Sasaki \cite{Nakano&Sasaki} carried out a
weak-field calculation of the self force in a Schwarzschild background by
evaluating the tail part of a Green's function. It was assumed, however,
that the correct force could be derived from a Green's function associated
with a certain Klein-Gordon type wave operator, instead of the harmonic
gauge-related Green's function as required in the MSTQW regularization scheme.
It is unclear yet whether or not the above regularization schemes produce,
in general, the results that would have been obtained by a strict
application of MSTQW regularization.

As we already mentioned, the calculation scheme to be presented in this
manuscript is based on MSTQW's formal expression for the gravitational
self force. It employs a technique---the {\em mode-sum regularization
scheme}---previously introduced \cite{Rapid,Barack} and tested
\cite{BurkoStatic,BurkoCircular,BB,BurkoShell,BurkoKerr} for the
scalar self force. In general, the application of this scheme consists of
two essentially independent parts: in the first part, one expands
the perturbation field of the particle into multipole modes (tensor harmonic
modes in the gravitational case, e.g.), and derives the ``bare'' self-force
$l$-modes associated with the $l$-modes of the metric perturbation.
The $l$-mode perturbation is finite and continuous even at the particle's
location, and the corresponding $l$-mode self-force maintains a finite
value as well (although it usually suffers a discontinuity across the worldline).
However, the sum over the bare force's $l$-modes turn out to be divergent.
In the second part of the mode-sum scheme, certain regularization parameters are
calculated analytically, by a local analysis of the Green's function
at large $l$ and small spacetime separations. These parameters are then used
to regularize the divergent sum over bare force's $l$-modes.
This calculation scheme completely relies on MSTQW's results, and
contains no further assumptions as to the appropriate regularization
(although it does contain certain assumptions concerning the mathematical
behaviour of various quantities involved in the analysis).

Previously, the above mode-sum scheme was fully implemented in
several test cases:
A scalar charge held static outside a Schwarzschild black hole
\cite{BurkoStatic}; A scalar charge in a circular orbit
\cite{BurkoCircular} or one in radial motion \cite{Barack,BB} in
Schwarzschild spacetime; and the motion of scalar or electric charges
on the background of a massive shell \cite{BurkoShell}. Recently,
Burko and Liu first applied the mode-sum scheme for a static scalar
charge in Kerr spacetime \cite{BurkoKerr} (however, an analytic foundation
for the scheme has not yet been established in the Kerr case).
So far, the mode-sum regularization scheme has never been applied for
calculating the gravitational self force.

The arrangement of this paper is as follows.
We start, in Sec.\ \ref{secII}, by summarizing and discussing
MSTQW's result for the regularized gravitational self force, which
is the basis for our calculation scheme. Specializing to the Schwarzschild
geometry, in Sec.\ \ref{secIII} we expand the (harmonic gauge) Green's
function in tensor harmonics, and obtain a set of equations for its various
multipole modes (i.e., a set which does not couple different $l,m$ modes,
although it does couple different tensor harmonic components).
In Sec.\ \ref{secIV} we then introduce our mode-sum scheme as applied to
the gravitational self-force problem. To demonstrate the applicability of
this scheme, in Sec.\ \ref{secV} we implement it in two simple cases by
explicitly calculating all necessary regularization parameters.
Sec.\ \ref{secVI} contains a summary of our results and a discussion
concerning the application of our scheme to other, more general cases.


\section{Worldline integral formulation of the gravitational self force}
\label{secII}

We consider a pointlike particle of mass $m$ moving in the exterior
of a black hole with mass $M\gg m$.\footnote{The extent to which the concept
of a pointlike particle makes sense in the context of the self-force
problem is discussed in Ref.\ \cite{QW}.} Let the metric $g_{\mu\nu}$ describe
the black hole background geometry, which we assume to be a solution of the vacuum
Einstein equation (later we specialize our discussion to the Schwarzschild
spacetimes). Let also $z^{\alpha}(\tau)$ represent the particle's trajectory,
and $u^{\alpha}(\tau)\equiv dz^{\alpha}/d\tau$ stand for its four-velocity.
The particle produces a small perturbation $h_{\alpha\beta}\ll g_{\alpha\beta}$
to the background geometry. The Einstein field equation for the metric
$g_{\alpha\beta}+h_{\alpha\beta}$, linearized in $h_{\alpha\beta}$, takes
the form
\begin{equation}\label{eqII10}
\Box \bar h_{\alpha\beta}(x)-2{{R^{\mu}}_{\alpha\beta}}\mbox{}^{\nu}\bar
h_{\mu\nu}(x)= -16\pi m \int_{-\infty}^{\infty}(-g)^{-1/2}\delta^4[x-z(\tau)]
u_{\alpha}(\tau)u_{\beta}(\tau) d\tau,
\end{equation}
where we have introduced the ``trace-reversed'' metric perturbation
\begin{equation}\label{eqII20}
\bar h_{\alpha\beta}=h_{\alpha\beta}-\frac{1}{2}g_{\alpha\beta}h
\end{equation}
(with $h\equiv g^{\mu\nu}h_{\mu\nu}$), and where we have set the
harmonic gauge condition $\bar h^{\mu\nu}{}_{;\nu}=0$.
In the perturbation equation (\ref{eqII10}), $\Box$ stands for the
covariant D'Alambertian operator,
$R_{\alpha\beta\gamma\delta}$ is the Riemann tensor in the background
geometry\footnote{We use here the convention of Ref.\ \cite{MTW} for the
Riemann tensor. Note the different convention used by Mino {\it et al.}
in \cite{MST}.}, $g$ stands for the determinant of the metric
$g_{\alpha\beta}$, and $\delta$ is the Dirac delta function.
The particle does not move along a geodesic of $g_{\alpha\beta}$, as it
interacts with its own field $h_{\alpha\beta}$. Phrased
differently, to the perturbation $h_{\alpha\beta}$ there corresponds a
self-force $F^{\alpha}$, in terms of which the particle's equation
of motion is given by
\begin{equation} \label{eqII30}
m a^{\alpha}(\tau)= F^{\alpha}(\tau),
\end{equation}
where $a^{\alpha}\equiv u^{\alpha}_{;\beta}u^{\beta}$ is the particle's
four-acceleration, with a semicolon denoting covariant differentiation
with respect to the {\em background} metric $g_{\alpha\beta}$.

Obviously, the perturbation $h_{\alpha\beta}$ diverges on the worldline
itself, and the ``bare'' self force associated with this perturbation
(as defined below) diverges as well. To obtain the physical equation of
motion, one must appropriately regularize the self force.
The analyses by MSTQW \cite{MST,QW} provide a formal expression for the
regularized self force, to order $O(m^2)$, in terms of a worldline integral over
derivatives of a retarded Green's function. It is found that, in a
vacuum background, $F^{\alpha}$ is solely due to the tail part of the Green's
function:
\begin{equation} \label{eqII40}
F^{\alpha}(\tau)=m^2 k^{\alpha\beta\gamma\delta}
\int_{-\infty}^{\tau^-}\bar G_{\beta\gamma\beta'\gamma';\delta}
[z^{\mu}(\tau);z^{\mu}(\tau')] u^{\beta'}(\tau')u^{\gamma'}(\tau')d\tau'.
\end{equation}
Here, the ``kinematic'' tensor $k^{\alpha\beta\gamma\delta}$ is given by
\begin{equation} \label{eqII50}
k^{\alpha\beta\gamma\delta}=\frac{1}{2}g^{\alpha\delta}u^{\beta}u^{\gamma}
        -g^{\alpha\beta}u^{\gamma}u^{\delta}
        -\frac{1}{2}u^{\alpha}u^{\beta}u^{\gamma}u^{\delta}
        +\frac{1}{4}u^{\alpha}g^{\beta\gamma}u^{\delta}
        +\frac{1}{4}g^{\alpha\delta}g^{\beta\gamma}
\end{equation}
(understood to be evaluated at the particle's location),
so designed to assure that the self-force has no component tangent
to the worldline (i.e., $F^{\alpha}u_{\alpha}=0$). This guarantees that
the mass $m$ maintains a constant value along the worldline.
The quantity $\bar G_{\beta\gamma\beta'\gamma'}$ is the two-point
retarded Green's function associated with the wave operator given
in the left-hand side of Eq.\ (\ref{eqII10}). It satisfies\footnote{
Note that although the wave operator defining the
Green's function $\bar G_{\alpha\beta\alpha'\beta'}$ indeed originates
from the perturbation equation in the harmonic gauge, the
Green's function itself does {\em not} satisfy the harmonic gauge
condition, as one can directly verify.
(To see this, note that the delta-function source does not
satisfy the conservation law---a vanishing covariant divergence---as does
the source for the metric perturbation itself. Consequently, the harmonic
gauge condition is not consistent with the Green's function equation as it
is with the perturbation equation.) Note that Eq.\ (2.12) in \cite{MST} is
therefore erroneous.}
\begin{eqnarray} \label{eqII60}
\Box \bar G_{\alpha\beta\alpha'\beta'}(x;x')
-2R^{\mu}{}_{\alpha\beta}{}^{\nu}(x)\bar G_{\mu\nu\alpha'\beta'}(x;x')
=
-16\pi(-g)^{-1/2}\bar g_{\alpha'(\alpha}(x,x')\bar g_{\beta)\beta'}(x,x')
\delta^4(x-x')\equiv Z_{\alpha\beta\alpha'\beta'},
\end{eqnarray}
with the supplementary causality condition
$\bar G_{\mu\nu\alpha'\beta'}(x;x')=0$
whenever $x$ lies outside the future light cone of $x'$. In this equation,
the D'Alambertian is taken with respect to $x$, parenthesized indices
indicate symmetrization, and $\bar g_{\alpha\alpha'}$ is the bi-vector of
geodesic parallel displacement defined in \cite{DB}.
In what follows we shall need only the value of this bi-vector in the
``coincidence'' limit:
$\lim_{x\to x'}\bar g_{\alpha\alpha'}= g_{\alpha\alpha'}$.
Note that the bi-tensor $\bar G_{\mu\nu\alpha'\beta'}(x;x')$ has, in general,
100 independent components (compared with 16 components in the
electromagnetic case, and only one in the scalar case).
We also mention that the trace-reversed metric perturbation itself
is constructed from the Green's function according to
\begin{equation} \label{eqII65}
\bar h_{\alpha\beta}(x)=m
\int_{-\infty}^{\infty}\bar G_{\alpha\beta\alpha'\beta'}
[x^{\mu};z^{\mu}(\tau)] u^{\alpha'}(\tau)u^{\beta'}(\tau)d\tau.
\end{equation}

To avoid confusion, it is worth commenting here about the different
notation previously used by different authors in writing Eq.\
(\ref{eqII40}):
The trace-reversed Green's function, denoted here by
$\bar G_{\alpha\beta\alpha'\beta'}$,
is denoted $G_{\mu\nu\alpha\beta}^{\rm ret}$ by MST \cite{MST},
$G_{aba'b'}^{\rm -}$ by QW \cite{QW,QWerror}, and
$\cal G_{\alpha\beta\alpha'\beta'}$ by Pfenning and Poisson \cite{EricNew}.
Note that Pfenning and Poisson use a different normalization for the
Green's function (it contains an extra factor $4$ with respect to either
\cite{MST,QW} and the normalization used here).
Also, note that in Ref.\ \cite{MST} the self-force is expressed
in terms of only the tail part of the Green's function, denoted therein
by $v_{\alpha\beta\alpha'\beta'}$, with no contribution from its
instantaneous part. This expression, however, is actually identical
to the above Eq.\ (\ref{eqII40}), as the worldline
integral involved is cutoff at $\tau^-$.

The Green's function $\bar G_{\mu\nu\alpha'\beta'}$ plays a central
role in our analysis. It is important to notice that, based on MSTQW
analysis, Eq.\ (\ref{eqII40}) is guaranteed to yield the correct,
``physical'' force, only when using the Green's function defined
through Eq.\ (\ref{eqII60}). The expression given in Eq.\ (\ref{eqII40})
may fail to represent the physical self force, and may even yield an
indefinite result, if a different Green's function is used.

For future use, it is useful to write the Green's function equation
(\ref{eqII60}) in the (non-covariant) form
\begin{equation}\label{eqII70}
\Box_s \bar G_{\alpha\beta\alpha'\beta'}+{\cal A}_{\alpha\beta}^{\mu\nu}
\bar G_{\mu\nu\alpha'\beta'}= Z_{\alpha\beta\alpha'\beta'}.
\end{equation}
Here, $\Box_s$ stands for the D'Alambertian operator acting on a scalar
function,
\begin{equation}\label{eqII80}
\Box_s \equiv \partial_{\alpha}\partial^{\alpha}
-g^{\alpha\beta}\Gamma^{\lambda}_{\alpha\beta}\partial_{\lambda}
\end{equation}
(where $\Gamma^{\lambda}_{\alpha\beta}$ are the connection coefficients),
and the ${\cal A}_{\alpha\beta}^{\mu\nu}$'s are certain differential
operators, of the first order at most, which describe how the various
components of the Green's function couple to each other. In a given
coordinate system we have
\begin{eqnarray} \label{eqII90}
{\cal A}_{\alpha\beta}^{\mu\nu} \bar G_{\mu\nu}=
2g^{\mu\nu}\left[
-2\Gamma^{\lambda}_{(\mu(\alpha} \bar G_{\lambda\beta),\nu)}
+\left(R^{\lambda}{}_{\mu(\beta\nu}-
\Gamma^{\lambda}_{\mu\nu,(\beta}+2\Gamma^{\rho}_{(\beta(\mu}
\Gamma^{\lambda}_{\nu)\rho}\right)\bar G_{\alpha)\lambda}
+\Gamma^{\lambda}_{\nu(\alpha}\Gamma^{\rho}_{\beta)\mu}\right]
-2R^{\mu}{}_{\alpha\beta}{}^{\nu}\bar G_{\mu\nu},
\end{eqnarray}
where the source-point indices $\alpha'\beta'$ have been suppressed
for brevity.
In the Appendix we give explicitly the operators
${\cal A}_{\alpha\beta}^{\mu\nu}$ for the Schwarzschild background case
(in Schwarzschild coordinates).


\section{Tensor Harmonic expansion and reduced equations} \label{secIII}

Accepting Eq.\ (\ref{eqII40}) as the basic expression for the gravitational
self force in curved
spacetime, the main concern remains how to implement this expression in
actual calculations. One may start by considering limiting cases, as the
weak field or slow motion scenarios, in which Eq.\ (\ref{eqII40}) could
be applied in a direct manner. This, indeed, was done recently by Pfenning
and Poisson in Ref.\ \cite{EricNew}.
However, in considering realistic black hole spacetimes, one ultimately wishes to
apply Eq.\ (\ref{eqII40}) for strong field self force calculations. Here, the
introduction of a mode-sum scheme seems inevitable. A mode-sum decomposition
approach is necessary for a dimensional reduction of the problem (in the usual
manner), but it is especially beneficial in the context of the self force
problem, because of the following reason: Whereas the metric perturbation
diverges on the worldline, its {\em individual modes}, as well as the
corresponding force modes, maintain a finite value even at the location of
the particle. Thus, in exploring the behavior of the individual self force
modes, one avoids dealing with divergent quantities.
Still, introducing a mode-sum scheme for calculating the self force is not a
straightforward task: the self force modes each carries a mixed imprint
of both the tail and instantaneous parts of the worldline integral, which
results in that the sum over modes usually turns out to diverge.
A carefully designed scheme for regularization of the mode sum is thus
necessary. The introduction of such a scheme is the main target of this
paper.

We first consider the multipole-mode decomposing of the Green's function.
As in the rest of this paper, we focus on the spherically symmetric
Schwarzschild black hole background, with a line element given by
\begin{equation} \label{eqIII10}
ds^2=-f(r)dt^2+f^{-1}(r)dr^2+r^2(d\theta^2+\sin^2\theta d\varphi^2),
\end{equation}
where $f(r)\equiv 1-2M/r$ and $M$ is the black hole's mass. Throughout this
paper we use Schwarzschild coordinates $t,r,\theta,\varphi$,
relativistic units (with $G=c=1$), and metric signature
${-}{+}{+}{+}$.

Any (sufficiently regular) symmetric covariant tensor of rank two,
$T_{\alpha\beta}$, can be expanded on a $2$-sphere in the form
\begin{equation} \label{eqIII15}
T_{\alpha\beta}=\sum_{l=0}^{\infty}\sum_{m=-l}^{l}\sum_{i=1}^{10}
T^{(i)lm}(r,t)Y^{(i)lm}_{\alpha\beta},
\end{equation}
where $T^{(i)lm}$ are scalar coefficients and $Y^{(i)lm}_{\alpha\beta}$
are the Regge-Wheeler-Zerilli tensor harmonics
\cite{RW,Zerilli1,Zerilli2,Thorne}.
In Schwarzschild coordinates $t,r,\theta,\varphi$, the set of tensor
harmonics $Y^{(i)lm}_{\alpha\beta}$ is given by\footnote{We adopt here
the orthogonal set introduced by Zerilli \cite{Zerilli2}, though we use a
different notation for the basis tensors: The symbols
$Y^{(1)lm}_{\alpha\beta}\ldots Y^{(10)lm}_{\alpha\beta}$ are used here
instead of Zerilli's
${\bf a}^{(0)}_{lm}$, ${\bf a}^{(1)}_{lm}$, ${\bf a}_{lm}$,
${\bf b}^{(0)}_{lm}$, ${\bf b}_{lm}$, ${\bf g}_{lm}$, ${\bf f}_{lm}$,
${\bf c}^{(0)}_{lm}$, ${\bf c}_{lm}$, and ${\bf d}_{lm}$, respectively.
Note the sign error in Eq.\ (A2j) of Ref.\ \cite{Zerilli2}.}
\begin{mathletters} \label{eqIII20}
\begin{equation} \label{eqIII20(1)}
Y^{(1)lm}_{\alpha\beta}=\left(
\begin{array}{c c c c}
1 & 0 & 0 & 0 \\
0 & 0 & 0 & 0 \\
0 & 0 & 0 & 0 \\
0 & 0 & 0 & 0
\end{array}\right) Y^{lm},
\end{equation}
\begin{equation} \label{eqIII20(2)}
Y^{(2)lm}_{\alpha\beta}=i/\sqrt{2}\left(
\begin{array}{c c c c}
0 & 1 & 0 & 0 \\
1 & 0 & 0 & 0 \\
0 & 0 & 0 & 0 \\
0 & 0 & 0 & 0
\end{array}\right) Y^{lm},
\end{equation}
\begin{equation} \label{eqIII20(3)}
Y^{(3)lm}_{\alpha\beta}=\left(
\begin{array}{c c c c}
0 & 0 & 0 & 0 \\
0 & 1 & 0 & 0 \\
0 & 0 & 0 & 0 \\
0 & 0 & 0 & 0
\end{array}\right) Y^{lm},
\end{equation}
\begin{equation} \label{eqIII20(4)}
Y^{(4)lm}_{\alpha\beta}=
\frac{ir}{\sqrt{2l(l+1)}}\left(
\begin{array}{c c c c}
0                 & 0 & \partial_{\theta} & \partial_{\varphi} \\
0                 & 0 &        0          &         0          \\
\partial_{\theta} & 0 &        0          &         0          \\
\partial_{\varphi}& 0 &        0          &         0
\end{array}\right) Y^{lm},
\end{equation}
\begin{equation} \label{eqIII20(5)}
Y^{(5)lm}_{\alpha\beta}=
\frac{r}{\sqrt{2l(l+1)}}\left(
\begin{array}{c c c c}
0 &        0           &        0          &          0         \\
0 &        0           & \partial_{\theta} & \partial_{\varphi} \\
0 & \partial_{\theta}  &        0          &          0         \\
0 & \partial_{\varphi} &        0          &          0
\end{array}\right) Y^{lm},
\end{equation}
\begin{equation} \label{eqIII20(6)}
Y^{(6)lm}_{\alpha\beta}=
r^2/\sqrt{2}\left(
\begin{array}{c c c c}
0 & 0 & 0 & 0 \\
0 & 0 & 0 & 0 \\
0 & 0 & 1 & 0 \\
0 & 0 & 0 & s^2
\end{array}\right) Y^{lm},
\end{equation}
\begin{equation} \label{eqIII20(7)}
Y^{(7)lm}_{\alpha\beta}=
\frac{r^2}{2\sqrt{\lambda l(l+1)}}\left(
\begin{array}{c c c c}
0 & 0 & 0   & 0        \\
0 & 0 & 0   & 0        \\
0 & 0 & D_2 & D_1       \\
0 & 0 & D_1 & -s^2 D_2
\end{array}\right) Y^{lm},
\end{equation}
\end{mathletters}
\begin{mathletters} \label{eqIII25}
\begin{equation} \label{eqIII25(8)}
Y^{(8)lm}_{\alpha\beta}=\frac{r}{\sqrt{2l(l+1)}}\left(
\begin{array}{c c c c}
0 &        0            & s^{-1}\partial_{\varphi} & -s\,\partial_{\theta} \\
0 &        0            &          0               &                       \\
s^{-1}\partial_{\varphi}&          0               &     0    &    0       \\
-s\,\partial_{\theta}   &          0               &     0    &    0
\end{array}\right) Y^{lm},
\end{equation}
\begin{equation} \label{eqIII25(9)}
Y^{(9)lm}_{\alpha\beta}=\frac{ir}{\sqrt{2l(l+1)}}\left(
\begin{array}{c c c c}
0 &        0                &        0                &           0         \\
0 &        0                & s^{-1}\partial_{\varphi} & -s\,\partial_{\theta} \\
0 & s^{-1}\partial_{\varphi}&        0                &          0          \\
0 & -s\,\partial_{\theta}   &        0                &          0
\end{array}\right) Y^{lm},
\end{equation}
\begin{equation} \label{eqIII25(10)}
Y^{(10)lm}_{\alpha\beta}=
\frac{-ir^2}{2\sqrt{\lambda l(l+1)}}\left(
\begin{array}{c c c c}
0 & 0 & 0          & 0            \\
0 & 0 & 0          & 0            \\
0 & 0 & -s^{-1}D_1 & s\,D_2       \\
0 & 0 & s\,D_2     & s\,D_1
\end{array}\right) Y^{lm},
\end{equation}
\end{mathletters}
where $Y^{lm}(\theta,\varphi)$ are the scalar spherical harmonics,
$s\equiv\sin\theta$, $\lambda \equiv (l-1)(l+2)/2$, and the operators
$D_1$ and $D_2$ are given by
\begin{eqnarray} \label{eqIII30}
D_1 \equiv 2(\partial_{\theta}-\cot\theta)\partial_{\varphi},
\quad\quad
D_2\equiv \partial_{\theta\theta}-\cot\theta\,\partial_{\theta}
-s^{-2} \partial_{\varphi\varphi}.
\end{eqnarray}
The ten\footnote{
Note that for $l=0$ and $l=1$ there are actually fewer independent tensor
harmonics: There are only $4$ independent harmonics at $l=0$ (the
harmonics $i=1,2,3$, and $6$), and $8$ independent harmonics for
each of the three $l=1$ modes (the harmonics $i=7$ and $10$ vanish
identically at $l=1$ for any value of the azimuthal number $m$).}
tensor harmonics of Eqs.\ (\ref{eqIII20}) and (\ref{eqIII25})
constitute an orthogonal set:
\begin{equation} \label{eqIII40}
\int d\Omega\, \eta^{\alpha\mu}\eta^{\beta\nu}[Y_{\mu\nu}^{(i)lm}]^*
Y_{\alpha\beta}^{(j)l'm'} =k^{(i)}\delta_{ij}\delta_{ll'}\delta_{mm'}
\quad\quad \text{(for $i,j=1,\ldots, 10$)},
\end{equation}
where $\eta^{\alpha\mu}\equiv{\rm diag}[-1,1,r^{-2},(rs)^{-2}]$,
an asterisk denotes complex conjugation, and the integration is carried
over a $2$-sphere (no summation over $i$ is implied on the right-hand
side). The coefficient $k^{(i)}$ takes the value $-1$ for
$i=2,4,8$, and $+1$ otherwise. The seven harmonics
$Y^{(1)lm}_{\alpha\beta}\ldots Y^{(7)lm}_{\alpha\beta}$ constitute a basis
for all symmetric covariant tensors of {\em even} parity, while the
remaining three harmonics  $Y^{(8)lm}_{\alpha\beta}\ldots
Y^{(10)lm}_{\alpha\beta}$ span all {\em odd} parity tensors. Recall that
the odd and even parity parts of a tensor (e.g., a metric perturbation)
are uncoupled, and can be treated separately \cite{RW}.
Using the orthogonality relation (\ref{eqIII40}), one easily
constructs the scalar coefficients of Eq.\ (\ref{eqIII15}) through
\begin{equation}\label{eqIII45}
T^{(i)lm}(r,t)=k^{(i)}\int d\Omega\,\eta^{\alpha\mu}\eta^{\beta\nu}
[Y_{\mu\nu}^{(i)lm}]^* T_{\alpha\beta}.
\end{equation}

Now, the Green's function $\bar G_{\alpha\beta\alpha'\beta'}(x,x')$ is a
bi-tensor. It transforms like a tensor at the evaluation point $x$
(when the coordinate transformation is carried out holding $x'$ fixed),
and it also transforms as a tensor at the source point $x'$
(when the transformation is performed with fixed $x$).
Regarding the Green's function, for a while, as a tensor at the evaluation
point $x$, we may expand it in tensor spherical harmonics at that point,
as in Eq.\ (\ref{eqIII15}). We write
\begin{equation}\label{eqIII50}
\bar G_{\alpha\beta\alpha'\beta'}(x,x')=
(rr')^{-1}\sum_{l,m,i}
\bar G^{(i)lm}_{\alpha'\beta'}(r,t;x')\,Y^{(i)lm}_{\alpha\beta}
\equiv \sum_{l}\bar G^{l}_{\alpha\beta\alpha'\beta'},
\end{equation}
where $\bar G^{(i)lm}_{\alpha'\beta'}$ are the multipole
expansion coefficients (independent of $\theta$ and $\varphi$),
the quantity $\bar G^{l}_{\alpha\beta\alpha'\beta'}$ is the one resulting
from formally summing over $i$ and $m$, and the radial factor
$(rr')^{-1}$ is introduced for later convenience.
We further need to expand the (bi-tensorial) source term of the Green's
function equation (\ref{eqII60}) in tensor spherical harmonics. This
expansion takes the form
\begin{equation}\label{eqIII60}
Z_{\alpha\beta\alpha'\beta'}=(rr')^{-1}\sum_{l,m,i}
Z^{(i)lm}_{\alpha'\beta'}(r,t;x')\,Y^{(i)lm}_{\alpha\beta},
\end{equation}
where, using Eq.\ (\ref{eqIII45}) with Eq.\ ({\ref{eqII60}), the
coefficients $Z^{(i)lm}_{\alpha'\beta'}$ are found to be given by
\begin{equation}\label{eqIII65}
Z^{(i)lm}_{\alpha'\beta'}=-16\pi k^{(i)} \delta(t-t')\delta(r-r')
g_{\rho'(\alpha'}(x') g_{\beta')\sigma'}(x')
\eta^{\rho'\mu'}(x')\eta^{\sigma'\nu'}(x') [Y^{(i)lm}_{\mu'\nu'}(\Omega')]^*
\end{equation}
(with $\Omega'$ standing for $\theta',\varphi'$).
Note here that the coefficients $Z^{(i)lm}_{\alpha'\beta'}$ depend on
the evaluation point $x$ only through the delta functions.\footnote{
In fact, the naive construction of the coefficients
$Z^{(i)lm}_{\alpha'\beta'}$ yields an expression involving
the bi-vector $\bar g_{\alpha\beta}$ [as in Eq.\ (\ref{eqII60})].
The form (\ref{eqIII65}) is then obtained by noticing that the
coefficients $Z^{(i)lm}_{\alpha'\beta'}$ transform like
scalars at the evaluation point $x$. This allows us to prime all tensorial
indices of the various factors involved in Eq.\ (\ref{eqIII65}) (i.e., take
these factors to transform like tensors with respect to the source point
$x'$), without affecting the value of $Z^{(i)lm}_{\alpha'\beta'}$.
The presence of the delta functions then further allows us to take all tensorial
factors in Eq.\ (\ref{eqIII65}) to be functions of only the source point
coordinates $x'$.}

Eq.\ (\ref{eqII70}) is now separable into multipole modes $l,m$ by
means of expansions (\ref{eqIII50}) and (\ref{eqIII60}):
By substituting these expansions in Eq.\ (\ref{eqII70}) one
obtains a set of equations which indeed couple between the ten functions
$\bar G^{(i=1\ldots 10)lm}_{\alpha'\beta'}$ (for given $l,m$),
but not between the different multipoles $l,m$.
To write the equations for the various multipole modes of the Green's
function, it is convenient to introduce the Eddington-Finkelstein
null coordinates
\begin{equation} \label{eqIII70}
v\equiv t+r^*, \quad u\equiv t-r^*\quad
\text{[where $dr^*/dr=f^{-1}(r)$],}
\end{equation}
and the time-radial operator
\begin{equation} \label{eqIII80}
D_{\rm s}^l\equiv \partial_{uv}+V_{\rm s}^l(r),
\end{equation}
where
\begin{equation} \label{eqIII90}
V_s^l(r)=\frac{f}{4}\left(\frac{f'}{r}+\frac{l(l+1)}{r^2}\right),
\end{equation}
with $f\equiv f(r)$ and $f'\equiv df(r)/dr$.
The operator $D_s^l$ is the familiar wave operator associated with the
$l$-mode of a massless scalar field in Schwarzschild spacetime
[note the relation
$r\Box_s[\phi(r,t)Y^{lm}(\Omega)/r]=-4f^{-1}(r)D_s^l\phi(r,t)Y^{lm}(\Omega)$,
where $\phi(r,t)$ is any function].
The equations for the various modes $\bar G^{(i)lm}_{\alpha'\beta'}$
then take the form
\begin{equation} \label{eqIII95}
D^l_{\rm s} \bar G^{(i)lm}_{\alpha'\beta'}+
{\cal I}^{(i)l}_{(j)}\bar G^{(j)lm}_{\alpha'\beta'}
=S_{\alpha'\beta'}^{(i)lm}\quad\quad \text{(sum over $j$)},
\end{equation}
where the source term is given by
\begin{equation}\label{eqIII110}
S_{\alpha'\beta'}^{(i)lm}=-[f(r)/4]Z_{\alpha'\beta'}^{(i)lm}=
8k^{(i)}\pi\delta(v-v')\delta(u-u')
g_{\rho'(\alpha'}(x') g_{\beta')\sigma'}(x')
\eta^{\rho'\mu'}(x')\eta^{\sigma'\nu'}(x')[Y_{\mu'\nu'}^{(i)lm}(\Omega')]^*,
\end{equation}
and where ${\cal I}^{(i)l}_{(j)}$ are differential operators that
couple between different $i$ values. These coupling terms read
(suppressing the indices $\alpha'\beta'$ and $lm$ for brevity)
\begin{mathletters}\label{eqIII100}
\begin{eqnarray}\label{eqIII100(1)}
{\cal I}^{(1)}_{(j)}\bar G^{(j)}=
\frac{1}{2}ff'\bar G^{(1)}_{,r}
-\frac{1}{8}\left(f'^2+4ff'/r\right) \bar G^{(1)}
+\frac{1}{8}\left(f^2f'^2-2f^3f''\right)\bar G^{(3)}
-\frac{1}{4}\sqrt{2}\,ff'\left(i\bar G^{(2)}_{,t}+r^{-1}f\bar
G^{(6)}\right),
\end{eqnarray}
\begin{eqnarray}\label{eqIII100(2)}
{\cal I}^{(2)}_{(j)}\bar G^{(j)}=
\frac{1}{4}(f'^2-ff''+2f^2/r^2)\bar G^{(2)}
+\frac{1}{4}\sqrt{2}\,iff'\left(f^{-2}\bar G^{(1)}_{,t}+
\bar G^{(3)}_{,t}\right)
-\frac{1}{2}\sqrt{l(l+1)}\,r^{-2}f \bar G^{(4)},
\end{eqnarray}
\begin{eqnarray}\label{eqIII100(3)}
{\cal I}^{(3)}_{(j)}\bar G^{(j)}=
-\frac{1}{2}ff'\bar G^{(3)}_{,r}
+\frac{1}{8}(8f^2/r^2-f'^2+4ff'/r)\bar G^{(3)}
+\frac{1}{8}(f^{-2}f'^2-2f^{-1}f'')\bar G^{(1)}
-\frac{1}{4}\sqrt{2}\,if^{-1}f'\bar G^{(2)}_{,t}
\nonumber\\
-\frac{1}{2}\sqrt{2}\,r^{-2}f\sqrt{l(l+1)}\,\bar G^{(5)}
+\frac{1}{4}\sqrt{2}\,(f'/r-2f/r^2)\bar G^{(6)},
\end{eqnarray}
\begin{eqnarray}\label{eqIII100(4)}
{\cal I}^{(4)}_{(j)}\bar G^{(j)}=
\frac{1}{4}ff'\left(\bar G^{(4)}_{,r}
-2r^{-1}\bar G^{(4)}+i\bar G^{(5)}_{,t}\right)
-\frac{1}{2}\sqrt{l(l+1)}\,(f^2/r^2)\bar G^{(2)},
\end{eqnarray}
\begin{eqnarray}\label{eqIII100(5)}
{\cal I}^{(5)}_{(j)}\bar G^{(j)}=
-\frac{1}{4}ff'\bar G^{(5)}_{,r}
-\frac{1}{4}f\left(f'/r-4f/r^2\right)\bar G^{(5)}
-\frac{1}{2}(f^2/r^2)\sqrt{2l(l+1)}\, \bar G^{(3)}
-\frac{1}{4}f^{-1}f'i\bar G^{(4)}_{,t} \nonumber\\
+\frac{1}{2}(f/r^2)\left[ \sqrt{l(l+1)}\, \bar G^{(6)}
-\sqrt{2\lambda}\, \bar G^{(7)}\right],
\end{eqnarray}
\begin{eqnarray}\label{eqIII100(6)}
{\cal I}^{(6)}_{(j)}\bar G^{(j)}=
\frac{1}{2}(f/r^2)(1-2rf')\bar G^{(6)}
-\frac{\sqrt{2}}{4}(f'/r)\bar G^{(1)}
-\frac{\sqrt{2}}{4}f^2(2f/r^2-f'/r)\bar G^{(3)}
+\frac{1}{2}\sqrt{l(l+1)}(f^2/r^2)\bar G^{(5)},
\end{eqnarray}
\begin{eqnarray}\label{eqIII100(7)}
{\cal I}^{(7)}_{(j)}\bar G^{(j)}=
-\frac{1}{2}(f/r^2)\bar G^{(7)}
-\frac{1}{2}(f^2/r^2)\sqrt{2\lambda}\,\bar G^{(5)},
\end{eqnarray}
\begin{eqnarray}\label{eqIII100(8)}
{\cal I}^{(8)}_{(j)}\bar G^{(j)}=
\frac{1}{4}ff'\left(\bar G^{(8)}_{,r}
-2r^{-1}\bar G^{(8)}-i\bar G^{(9)}_{,t}\right)
\end{eqnarray}
\begin{eqnarray}\label{eqIII100(9)}
{\cal I}^{(9)}_{(j)}\bar G^{(j)}=
-\frac{1}{4}ff'\bar G^{(9)}_{,r}
-\frac{1}{4}f\left(f'/r-4f/r^2\right)\bar G^{(9)}
+\frac{1}{4}f^{-1}f'i\bar G^{(8)}_{,t}
-\frac{1}{2}(f/r^2)\sqrt{2\lambda}\, \bar G^{(10)},
\end{eqnarray}
\begin{eqnarray}\label{eqIII100(10)}
{\cal I}^{(10)}_{(j)}\bar G^{(j)}=
-\frac{1}{2}(f/r^2)\bar G^{(10)}
-\frac{1}{2}(f^2/r^2)\sqrt{2\lambda}\,\bar G^{(9)}.
\end{eqnarray}
\end{mathletters}
The separated equations (\ref{eqIII95}) have the convenient property
that no coupling between the various modes occur in the main
parts of the equations (i.e., the parts containing second derivatives).
Coupling between different $i$-modes comes into action only through the
${\cal I}^{(i)}_{(j)}$ terms, which contain one or no derivatives.
Note that the even parity modes ($i=1\ldots 7$) do not couple at all
to the odd parity modes ($i=8\ldots 10$). The two mode types form a
disjoint set of equations, as one would expect.


\section{Mode sum regularization} \label{secIV}

Without loss of generality, let us take the point along the particle's
trajectory where we wish to calculate the self force to be at $\tau=0$.
Based on Eq.\ (\ref{eqII40}), we may express the self force as
\begin{eqnarray}\label{eqIV3}
F_{\alpha}= F_{\alpha}^{\rm (bare)} - F_{\alpha}^{\rm (inst)},
\end{eqnarray}
where
\begin{equation}\label{eqIV5}
F_{\alpha}^{\rm (bare)}=
m^2 k_{\alpha}{}^{\beta\gamma\delta}
\int_{-\infty}^{0^+}\bar G_{\beta\gamma\beta'\gamma';\delta}
[z^{\mu}(\tau=0);z^{\mu}(\tau')] u^{\beta'}(\tau')u^{\gamma'}(\tau')d\tau'=
m k_{\alpha}{}^{\beta\gamma\delta}\bar h_{\beta\gamma;\delta}
\end{equation}
is the ``bare'' force associated with the metric
perturbation $\bar h_{\alpha\beta}$ [the second equality here stems from
Eq.\ (\ref{eqII65})], and
\begin{equation}\label{eqIV10}
F_{\alpha}^{\rm (inst)}=
\lim_{\epsilon\to 0^+} \delta F^{(\epsilon)}_{\alpha}\equiv
\lim_{\epsilon\to 0^+} \left[ m^2 k_{\alpha}{}^{\beta\gamma\delta}
\int_{-\epsilon}^{0^+}\bar G_{\beta\gamma\beta'\gamma';\delta}
[z^{\mu}(\tau=0);z^{\mu}(\tau')]u^{\beta'}(\tau')u^{\gamma'}(\tau')d\tau'\right]
\end{equation}
is the ``instantaneous'' part of the force.
The quantities $F_{\alpha}^{\rm (bare)}$ and
$F_{\alpha}^{\rm(inst)}$---both involving integration through the particle's
location---are singular and so poorly defined as they stand.
For definiteness, we may redefine the integrands appearing in Eqs.\
(\ref{eqIV5}) and (\ref{eqIV10}) as vector fields in the
neighborhood of the evaluation point, and later be interested in
their behavior on the worldline. Note, however, that the difference
$F_{\alpha}^{\rm (bare)}-F_{\alpha}^{\rm(inst)}$ {\em does} yield a definite
finite value at the particle's location: According to the analysis by MSTQW,
this value represents the physical self-force $F_{\alpha}$.

To introduce our mode-sum regularization scheme, let us denote by
$F_{\alpha}^{l{\rm (bare)}}$ and $\delta F_{\alpha}^{(\epsilon)l}$,
respectively, the contributions to $F_{\alpha}^{{\rm (bare)}}$ and
$\delta F_{\alpha}^{(\epsilon)}$ coming from the $l$-mode of the Green's
function [these two quantities are obtained by replacing the Green's
function $\bar G_{\beta\gamma\beta'\gamma'}$ in Eqs.\ (\ref{eqIV5}) and
(\ref{eqIV10}) with its $l$-mode $\bar G^{l}_{\alpha\beta\alpha'\beta'}$
defined in Eq.\ (\ref{eqIII50})]. We may then express the self force as a
sum over $l$-modes, in the form\footnote{
It is assumed here that the differentiation and the integration
involved in constructing $F_{\alpha}^{{\rm (bare)}}$ and
$\delta F_{\alpha}^{(\epsilon)}$ out of
$\bar G_{\beta\gamma\beta'\gamma'}$ can both be performed term by term
with respect to the sum over $l$.}
\begin{equation}\label{eqIV50}
F_{\alpha}=
\lim_{\epsilon\to 0^+}\sum_{l=0}^{\infty}
\left(F^{l{\rm (bare)}}_{\alpha}-\delta F^{(\epsilon)l}_{\alpha} \right).
\end{equation}
An essential feature of our scheme arises from the fact that the $l$-mode
of the metric perturbation $h_{\alpha\beta}$ is everywhere finite and
continuous; it remains finite and continuous even at the location of the
particle, where the overall perturbation diverges. Consequently,
the $l$-modes of the bare and instantaneous forces,
$F_{\alpha}^{l{\rm (bare)}}$ and $\delta F_{\alpha}^{(\epsilon)l}$,
turn out to attain finite values. This behavior has been analyzed
and demonstrated in the analogous scalar self force problem
\cite{Rapid,Barack}, and is equally valid in the gravitational case
as well (we demonstrate this in the next section).
As in the scalar self-force model, the two quantities $F^{l{\rm (bare)}}_{\alpha}$
and $\delta F^{(\epsilon)l}_{\alpha}$ are discontinuous through the particle's
location [regarding the integrands appearing in Eqs.\ (\ref{eqIV5}) and
(\ref{eqIV10}) as vector fields in the neighborhood of the particle]. That is,
each of these two quantities attains different (finite) values if calculated by
taking the limit $r\to (z^r)^+$ or else the limit $r\to (z^r)^-$.
[Later we assign to $F^{l{\rm (bare)}}_{\alpha}$ and $\delta F^{(\epsilon)l}_{\alpha}$
the labels $+$ or ${-}$ to indicate weather they were calculated
from $r\to (z^r)^+$ or rather from $r\to (z^r)^-$, respectively.]
Note, however, that the difference $F^{l{\rm (bare)}}_{\alpha}-
\delta F^{(\epsilon)l}_{\alpha}$ [as well as the sum over modes in
Eq.\ (\ref{eqIV50}), producing the regularized force $F_{\alpha}$] does not
depend on the direction from which the limit is taken.

Although each of the bare modes $F^{l{\rm (bare)}}_{\alpha}$ yields
a finite contribution to the self force, the infinite sum over
$F^{l{\rm (bare)}}_{\alpha}$ diverges, in general. This is easily
demonstrated already in the simple case of a static mass in flat space,
a case studied in the following section (see also \cite{Barack} for a
discussion of this point in the analogous scalar case).
To carry out the regularization procedure, one seeks an
($\epsilon$-independent) function $H^{l\pm}_{\alpha}$, such that the
series $\sum_{l}(F^{l{\rm (bare)}\pm}_{\alpha} - H^{l\pm}_{\alpha})$ would
converge.
Once such a function is found, Eq.\ (\ref{eqIV50}) can be written as
\begin{equation}\label{eqIV60}
F_{\alpha}=\sum_{l=0}^{\infty}\left(F^{l{\rm (bare)}\pm}_{\alpha}
-H^{l\pm}_{\alpha} \right) - D^{\pm}_{\alpha},
\end{equation}
where
\begin{equation}\label{eqIV70}
D^{\pm}_{\alpha} \equiv \lim_{\epsilon\to 0^+}\sum_{l=0}^{\infty}
\left(\delta F^{(\epsilon)l\pm}_{\alpha} - H^{l\pm}_{\alpha}\right).
\end{equation}
The parameter $D^{\pm}_{\alpha}$ is well defined, since, by construction of
$H^{l\pm}_{\alpha}$, the difference
$\delta F^{(\epsilon)l\pm}_{\alpha} - H^{l\pm}_{\alpha}$ yields a convergent
sum over $l$ [note that the two quantities $F^{l{\rm (bare)}\pm}_{\alpha}$
and $\delta F^{(\epsilon)l\pm}_{\alpha}$ must bare the same singular
behavior at large $l$, as their difference yields a convergent sum over
$l$---see Eq.\ (\ref{eqIV50})]. Also note here that the limit
$\epsilon \to 0$ and the sum over $l$ should {\em not} be interchanged;
otherwise, the crucial contribution from $\delta F^{(\epsilon)l\pm}_{\alpha}$
would be lost.

Eq.\ (\ref{eqIV60}) constitutes the basic expression for the gravitational
self force in the framework of the mode sum regularization approach.
The implementation of this expression for calculating the self force at a
certain point along a given trajectory involves two (essentially
independent) parts: In the first part, one should first calculate (numerically,
in general) the $l$-modes of the bare metric perturbation (in the harmonic
gauge) and then use these modes to construct the bare force modes $F^{l{\rm
(bare)}}_{\alpha}$.
In the second part of the calculation procedure one should obtain the
``regularization function'' $H^l_{\alpha}$.
In principle, this function should be constructed by exploring the
asymptotic behavior of the bare modes $F^{l}_{\alpha}$ as $l \to \infty$.
It is more convenient, however, to read this large $l$ asymptotic behavior
from the quantity $\delta F^{(\epsilon)l}_{\alpha}$, which is strictly
local (recall that $F^{l{\rm (bare)}}_{\alpha}$
and $\delta F^{(\epsilon)l}_{\alpha}$ bare the same singular
behavior at large $l$).
To this end, one merely needs the asymptotic behavior of
$\delta F^{(\epsilon)l}_{\alpha}$ in the immediate neighborhood of
$\epsilon=0$. This allows one to derive $H^{l}_{\alpha}$ (and later
also $D_{\alpha}$) using local analytic methods, as we shall
demonstrate in the next section.

In general, the (one-sided values of the) bare modes
$F^{l{\rm (bare)}}_{\alpha}$ and $\delta F^{(\epsilon)l}_{\alpha}$
are found to diverge at large $l$ as $\propto l$. This was demonstrated
for various trajectories in the scalar self force problem, and will be
demonstrated below in the gravitational case as well.
To construct the regularization function $H^{l}_{\alpha}$ so as
to regularize the mode-sum in Eq.\ (\ref{eqIV60}), one should
therefore derive the three leading-order terms in the $1/l$ expansion
of $\delta F^{(\epsilon)l}_{\alpha}$. It appears more convenient
to use an expansion in powers of the new variable
\begin{equation}\label{eqIV80}
L\equiv l+1/2.
\end{equation}
Denoting the coefficients of the leading-order terms
by $A_{\alpha}$, $B_{\alpha}$, and $C_{\alpha}$, we shall have, in
general,
\begin{equation}\label{eqIV90}
H^{l}_{\alpha}=A_{\alpha}L+B_{\alpha}+C_{\alpha}/L.
\end{equation}
The implementation of the mode sum regularization scheme therefore amounts to
(i) Calculating the bare modes $F^{l{\rm (bare)}}_{\alpha}$ (this
is usually done numerically); (ii) Deriving the four
``regularization parameters'' $A_{\alpha}$, $B_{\alpha}$, $C_{\alpha}$,
and $D_{\alpha}$ (by local analytic methods); and (iii) summing
over $l$ using Eq.\ (\ref{eqIV60}) to obtain the regularized force
$F_{\alpha}$.

The scheme described here is based on the prescription (\ref{eqII40}),
which is formulated within the harmonic gauge. It therefore requires
that the bare force modes be obtained from the metric perturbation in the
harmonic gauge. This poses a problem from a practical point of view,
as a separable formalism for the metric perturbation in the harmonic
gauge has not been well developed as it has for other gauges [especially
the Regge-Wheeler (RW) gauge].\footnote{This problem becomes more acute when
dealing with Kerr spacetime, for which a separation formalism for the metric
perturbation has been developed so far only in the radiation gauge
\cite{Chrzanowski}.}
We shall deal with this gauge problem in a forthcoming paper \cite{gauge},
where an attempt is made to rephrase the scheme in terms of other gauges,
such as the RW gauge.

Before proceeding to discuss some simple applications of the method
proposed here, we should comment on a certain technical
subtlety which we avoided so far in our discussion.
As we thoroughly discuss in Ref.\ \cite{Barack} (in the context of the
scalar self force problem), the Green's function does not admit a
convergent multipole expansion, as a result of its being singular along the
light cone of the source point. As a consequence, the modes
$F^{l{\rm (bare)}}_{\alpha}$ (and $\delta F^{(\epsilon)l}_{\alpha}$)
contain certain terms which oscillate rapidly at large $l$, rendering
the sums over $l$ in Eqs.\ (\ref{eqIV50}), (\ref{eqIV60}), and
(\ref{eqIV70}) non-convergent. In Ref.\ \cite{Barack} we justified
throwing away these divergent oscillatory terms. To formalize the
omission of these terms, we therein introduced a new summation and limit
operations (the ``tilde-summation'' and ``tilde-limit'') which eliminate
any oscillatory divergent terms while preserving the monotonic behaviour.
The same problem---with the same solution---persists in our current
gravitational case. However, to avoid complexity in our current
presentation, we shall not attempt to indicate explicitly where a
tilde-operation is to be applied [as in above Eqs.\ (\ref{eqIV50}),
(\ref{eqIV60}), and (\ref{eqIV70})]. In the analysis to follow, a
tilde-summation or a tilde-limit will be implicitly used when appropriate.


\section{Simple applications}\label{secV}

We now demonstrate the applicability of the above calculation scheme
in two simple test cases. First, we consider the trivial case of a static
point mass in flat spacetime. This would provide a simple test
case (where the result is obvious: a vanishing self force)
against which we may check the validity of our scheme.
We then move on to the Schwarzschild spacetime, and
consider a freely falling particle on a radial geodesic.
In both cases we construct all four necessary regularization parameters.
For simplicity, when considering the second case we shall focus on
calculating the force at a presumed turning point of the geodesic
(i.e., where $dz^r/d\tau=0$), for which case the calculation becomes
considerably more simple (see below). We emphasize that our calculation
and results apply equally well for either weak or strong field.
The application of the scheme in more realistic cases (ones of
greater astrophysical relevance) will be presented elsewhere \cite{BL}.

\subsection{Static particle in flat space}

We consider a static particle of mass $m$ in Minkowski spacetime. The particle
is located at $r=r_0$, in a certain spherical coordinate system
$t,r,\theta,\varphi$. (We adopt here spherical coordinates in order to make
the calculation more closely related to the Schwarzschild case discussed
later.) The four-velocity of this static particle is
$u^{\alpha}(\tau)= \delta^{\alpha}_{t}$ at all $\tau$, and thus
only the Green's function components $\bar G_{\beta\gamma t't'}$ take part in
constructing the force through Eq.\ (\ref{eqII40}).
Considering the system of equations (\ref{eqII70}) for the $10$
functions $\bar G_{\beta\gamma t't'}$, one finds that the source
$Z_{\alpha\beta t't'}$ is nonvanishing only for $\alpha\beta=tt$.
[To see this, note in Eq.\ (\ref{eqIII65}) that for $\alpha'\beta'=t't'$
the only contribution comes at $i=1$. Then, the only contribution
to $Y^{(1)lm}_{\alpha\beta}$ in Eq.\ (\ref{eqIII60}) is at $\alpha\beta=tt$.]
Since all interaction terms
${\cal A}^{tt}_{\alpha\beta}\bar G_{ttt't'}$ appearing
in Eqs.\ (\ref{eqII70}) vanish in Minkowski spacetime [see Eqs.\
(\ref{App1a})--(\ref{App1j}), with $f'=0$], we find that only the
sourced component $G_{ttt't'}$ takes a nonzero value, while all other
components $\bar G_{\beta\gamma t't'}$ (which satisfy homogeneous equations)
vanish. Thus, for $\alpha'\beta'=t't'$, the system of equations
(\ref{eqII70}) reduces to a single equation for the quantity
$\bar G_{ttt't'}\equiv \bar G$:
\begin{equation}\label{eqV10}
\Box_s \bar G = Z_{ttt't'}=-16\pi(-g)^{-1/2}\delta^4(x-x').
\end{equation}
Note that in the simple case considered here---a static particle in flat
spacetime---only {\em one} Green's function component out of $100$ actually
takes part in the computation of the self force.

The self force can now be constructed from Eq.\ (\ref{eqII40}) by setting
$u^{\alpha}=\delta^{\alpha}_{t}$ and $u^{\alpha'}=\delta^{\alpha'}_{t'}$,
and recalling that all component $G_{\beta\gamma t't'}$ but
$G_{ttt't'}$ vanish. We find (evaluating the force at $\tau=0$ without
loss of generality)
\begin{equation} \label{eqV20}
F_{\alpha}= \frac{1}{4}m^2 \int_{-\infty}^{0^-}
\bar G_{,\alpha}[z^{\mu}(\tau=0);z^{\mu}(\tau')] d\tau' \quad\quad
\text{for $\alpha=r,\theta,\varphi$},
\end{equation}
as well as $F_{t}=0$. [For $\alpha=t$, the integrand in Eq.\ (\ref{eqII40})
in identically zero. This is a trivial result for a static particle,
as the force is known to satisfy the normalization condition,
$F_{\alpha}u^{\alpha}=0$.]

The Green's function equation (\ref{eqV10}) is exactly the same as
the Green's function equation for a scalar field [compare with
Eq.\ (4) in Ref.\ \cite{Barack}], apart from a relative numerical
factor of $4$ on the right-hand side (the source in the gravitational
case is greater by $4$ than the source in the scalar case). In addition,
the construction of the self force from the Green's function through Eq.\
(\ref{eqV20}) is exactly the same as the construction of the tail part
of the force in the scalar case [compare with Eqs.\ (12) and (13) of Ref.\
\cite{Barack}], apart from a relative factor of $1/4$
on the right-hand side, which compensates for the extra factor of $4$
in the Green's function equation. We may conclude, in particular, that
the bare modes of the self force acting on our static particle of mass $m$
are equal to the bare modes of the scalar self force acting on a static particle
of scalar charge $q=m$. It is then possible to simply use the results
already obtained in the scalar case: For the regularization parameters
we obtain (taking the Minkowski limit of the results described in Refs.\
\cite{Rapid,Barack} and replacing $q\to m$),
\begin{equation}\label{eqV30}
A_{\alpha}^{\pm}=\mp \frac{m^2}{r_0^2}\,\delta_{\alpha}^{r},
\quad\quad
B_{\alpha}=-\frac{m^2}{2r_0^2}\,\delta_{\alpha}^{r},
\quad\quad
C_{\alpha}=D_{\alpha}=0.
\end{equation}
As the average of the two sided values of $A_{\alpha}$ vanishes, we
finally obtain from Eq.\ (\ref{eqIV60})
\begin{equation}\label{eqV40}
F_{\alpha}=\sum_{l=0}^{\infty}\left(\bar F^{l{\rm (bare)}}_{\alpha}
-B_{\alpha} \right),
\end{equation}
where
$\bar F^{l{\rm (bare)}}\equiv [F^{l{\rm (bare)}+}+F^{l{\rm (bare)}-}]/2$.

Now, in our current trivial case, the averaged bare-force modes appearing in
Eq.\ (\ref{eqV40}) are easily calculated \cite{Barack}: Solving first for
the metric perturbation modes\footnote{
Interestingly, in our case---a static particle in flat spacetime---the
metric perturbation in the harmonic gauge exactly coincides with the one in
the RW gauge.}, one finds that all components of the trace-reversed
modes $\bar h_{\alpha\beta}^l$ vanish, except $\bar h_{tt}^l$,
which is given by $\bar h_{tt}^l=mr^{-l-1}r_0^l$ (for $r>r_0$) and
$\bar h_{tt}^l=mr^lr_0^{-l-1}$ (for $r<r_0$). Then, using the second
equality of Eq.\ (\ref{eqIV5}), one obtains $F_r^{l+}=-(l+1)m^2r_0^{-2}$ and
$F_r^{l-}=lm^2r_0^{-2}$, yielding $\bar F_r^l=-m^2/(2r_0^2)$ (with all other
components vanishing). Thus, the averaged bare modes of the force are
found to be $l$-independent, each identically equal to the regularization
parameter $B_{\alpha}$. Consequently, one finds that each of the terms in the
sum over $l$ in Eq.\ (\ref{eqV40}) vanishes independently, with an obvious
(and expected) vanishing of the overall self force.

\subsection{Radial geodesic motion in Schwarzschild, at a turning point}

Let us now consider a particle moving in a radial geodesic in
Schwarzschild spacetime. For this orbit we have, identically,
$u^{\theta}=u^{\varphi}=0$, and therefore only the Green's function components
$\bar G_{\alpha\beta t't'}$, $\bar G_{\alpha\beta t'r'}=\bar G_{\alpha\beta r't'}$,
and $\bar G_{\alpha\beta r'r'}$ would take part in constructing the force
through Eq.\ (\ref{eqII40}).
Now, for any given combination of $\alpha'\beta'$, Eq.\ (\ref{eqII70})
constitutes a set of $10$ coupled equations for the $10$ independent
quantities $\bar G_{\alpha\beta\alpha'\beta'}$. Considering the three sets
of equations (\ref{eqII70}) with $\alpha'\beta'=t't'$, $t'r'$, and
$r'r'$, we find that the source $Z_{\alpha\beta\alpha'\beta'}$ is
nonvanishing only for $\alpha\beta=\alpha'\beta'$ (e.g., in the set
of equations for $\bar G_{\alpha\beta t't'}$, only the equation for
$\bar G_{ttt't'}$ is sourced).
Similarly, Eq.\ (\ref{eqIII95}) forms, for any specific value of
$\alpha'\beta'$, a set of ten equations for the ten tensor-harmonic
modes $\bar G^{(i=1\ldots 10)lm}_{\alpha'\beta'}$. The source
$S^{(i)lm}_{\alpha'\beta'}$ for these equations is nonvanishing
only at $i=1$ for $\alpha'\beta'=t't'$, at $i=2$ for
$\alpha'\beta'=t'r',r't'$, or at $i=3$ for $\alpha'\beta'=r'r'$.
Clearly, the three odd-parity modes $i=8,9,10$, which are not sourced
and also do not couple to any of the even parity modes $i=1,2,3$,
would all vanish (this is expected, of course, by virtue of our physical
setup, which only includes an even-parity source).
However, the four even-parity modes $i=4,5,6,7$, although not sourced
in Eqs.\ (\ref{eqIII95}), do couple to the modes
$i=1,2,3$ and will therefore not vanish, in general.

In conclusion of the above discussion, we find that in the scenario
considered here---a particle moving radially on a spherically
symmetric background---one has to deal with $3$ sets of $7$ coupled
equations for a total of $21$ nontrivial independent components
$\bar G^{(i)lm}_{\alpha'\beta'}$: One of these sets (corresponding to
$\alpha'\beta'=t't'$) contains a source term only at $i=1$, the other
set (corresponding to $\alpha'\beta'=t'r'$ or $r't'$) is sourced only
at $i=2$, and the third (for $\alpha'\beta'=r'r'$) only at $i=3$.
To write these three sets of equations in a
simple form, it is convenient to re-define our spherical coordinate
system, such that the radial trajectory would be directed along the
polar axis. In this spherical system, the Green's function (now sourced
only at $\theta'=0$) would contain only axially-symmetric, $m=0$ modes.
We then also introduce the new variables $\tilde G^{(i)l}_{\alpha'\beta'}$,
defined (for $\alpha'\beta'=t't',t'r',r't'$, or $r'r'$) through
\begin{equation} \label{eqV47}
\bar G^{(i)l}_{\alpha'\beta'}\equiv  8\pi
g_{\rho'(\alpha'}(x') g_{\beta')\sigma'}(x')\eta^{\rho'\mu'}\eta^{\sigma'\nu'}
a^{(i)}Y^{l}(\theta')\tilde G^{(i)l}_{\mu'\nu'}
\quad\text{(no summation over $i$)},
\end{equation}
where $Y^{l}(\theta)\equiv Y^{l,m=0}$ and $a^{(i)}=1$ for all $i$, except
$a^{(2)}=i/\sqrt{2}$.
Then, each of the above three sets of equations (for $\alpha'\beta'=t't'$,
$t'r'$, or $r'r'$) takes the form
\begin{equation} \label{eqV45}
D^l_{\rm s} \tilde G^{(i)l}_{\alpha'\beta'}+
\tilde{\cal I}^{(i)l}_{(j)}\tilde G^{(j)l}_{\alpha'\beta'}
=q^{(i)}_{\alpha'\beta'}\delta(u-u')\delta(v-v'),
\end{equation}
where
\begin{equation} \label{eqV49}
q^{(i)}_{\alpha'\beta'}=\left\{
\begin{array}{ll}
\delta_{\alpha'}^{t'}\delta_{\beta'}^{t'},   &  i=1, \\
2\delta_{(\alpha'}^{t'}\delta_{\beta')}^{r'}, &  i=2, \\
\delta_{\alpha'}^{r'}\delta_{\beta'}^{r'},   &  i=3, \\
0,                                           &  i=4,\ldots,7,
\end{array}\right.
\end{equation}
and $\tilde{\cal I}^{(i)l}_{(j)}={\cal I}^{(i)l}_{(j)}$ for all $i,j$,
except $\tilde{\cal I}^{(i\ne 2)l}_{(j=2)}=a^{(2)}{\cal I}^{(i)l}_{(2)}$
and $\tilde{\cal I}^{(i=2)l}_{(j\ne 2)}={\cal I}^{(2)l}_{(j)}/a^{(2)}$.
Finally, to express the $l$-mode Green's function $\bar G^l_{\alpha\beta\alpha'\beta'}$
in terms of the new variables $\tilde G^{(i)l}_{\alpha'\beta'}$, we
substitute for $\bar G^{(i)l}_{\alpha'\beta'}$ in Eq.\ (\ref{eqIII50}), and
consider only the Green's function components with $\alpha\beta=tt,tr,rr$
(we shall need only these three components in our following analysis).
Recalling $Y^{l}(\theta)=[(2l+1)/(4\pi)]^{1/2}P_l(\cos\theta)=
[L/(2\pi)]^{1/2}P_l(\cos\theta)$ (where $P_l$ is the Legendre polynomial)
and $P_l(\cos\theta')=1$ for $\theta'=0$, we then obtain for the three
relevant components of the $l$-mode Green's function, evaluated at the
polar axis ($\theta=0$),
\begin{equation}\label{eqV60}
\bar G^{l}_{tt\alpha'\beta'}= 4(rr')^{-1}f^2(r')
L\,\tilde G^{(1)l}_{\alpha'\beta'},   \quad\quad
\bar G^{l}_{tr\alpha'\beta'}= -2(rr')^{-1}
L\,\tilde G^{(2)l}_{\alpha'\beta'},   \quad\quad
\bar G^{l}_{rr\alpha'\beta'}= 4(rr')^{-1}f^{-2}(r')
L\,\tilde G^{(3)l}_{\alpha'\beta'}.
\end{equation}
[We may already, at this stage, evaluate the Green's function at the
polar axis (where the particle is located), as the following construction
of the $r$ component of the force does not involve differentiation of the
Green's function with respect to neither $\theta$ nor $\varphi$---see Eq.\
(\ref{eqV95}) below.]

To implement our mode sum prescription, we would now like to derive the
regularization parameters $A_{\alpha}$, $B_{\alpha}$, $C_{\alpha}$
and $D_{\alpha}$. This task turns out to be considerably simpler
(though not trivial) when the self force is evaluated at a presumed
turning point of the radial geodesic.
As a first non-trivial demonstration of applying our mode-sum regularization
scheme, we hereafter focus on this special case.
That is, we assume that the geodesic particle is momentarily at rest
(i.e., $u^r=0$) at $r= r_0$, and calculate the local self force at that
point. For simplicity, we shall consider in our calculation only the
$r$ component of the force.\footnote {The angular components $F_{\theta}$ and
$F_{\varphi}$ are expected to vanish due to the symmetry of our physical
setup, although we shall not attempt here to verify that our scheme indeed
leads to this obvious result. At a turning point we also have $F_t=0$,
stemming, in a trivial manner, from the orthogonality relation
$F_{\alpha}u^{\alpha}=0$.}
Extension of this calculation to an arbitrary point of a radial geodesic
is straightforward though rather laborious, and will be treated
elsewhere \cite{BL}.
On the technical side, the calculation of the regularization
parameters very much resembles the calculation in the scalar case,
which we describe in much detail in Ref.\ \cite{Barack}. For this reason,
we avoid here many of the technicalities involved in deriving the
parameters, and refer the reader to Ref.\ \cite{Barack} for more details.
In what follows we only give a very general description of how
the analysis proceeds in our current gravitational case.

As explained in Sec.\ \ref{secIV}, the mode-sum scheme's regularization parameters
can be derived by exploring the behavior of the quantity
$\delta F^{(\epsilon)l}_{\alpha}$ [see Eq.\ (\ref{eqIV10})] at small
$\epsilon$, using a perturbation analysis of the Green's function
$l$-modes at large-$l$ and small spacetime deviations. In our current
problem of a radially moving particle at a turning point, the radial
component $\delta F^{(\epsilon)l}_r$ is constructed from the various
components of the $l$-mode Green's function by
\begin{equation} \label{eqV95}
\delta F^{(\epsilon)l}_r=\frac{1}{4}m^2 f^{-1} \int_{-\epsilon}^{0^+}
\left[\bar G^l_{tt\alpha'\beta',r}+f'f^{-1}\bar G^l_{tt\alpha'\beta'}
-4\bar G^l_{rt\alpha'\beta',t}+f^2\bar G^l_{rr\alpha'\beta',r}+
3ff'\bar G^l_{rr\alpha'\beta'}\right]u^{\alpha'}(\tau')u^{\beta'}(\tau')
d\tau'
\end{equation}
(with $f$ and $f'\equiv df/dr$ understood to be evaluated at $r_0$), which is
obtained from Eq.\ (\ref{eqIV10}) by setting $u^r=u^{\theta}=u^{\varphi}=0$
and $u^t=f^{-1/2}$.

To analyze the large $l$ behavior of the $l$-modes
$\bar G^l_{\alpha\beta\alpha'\beta'}$ appearing in Eq.\ (\ref{eqV95}),
we first introduce the ``neutralized'' spacetime deviation
variables\footnote
{The variables $\Delta_{r }$, $\Delta_{r'}$, $\Delta_{t }$, and $z$ are
``neutral'' in the sense that they each consist of a small, $O(\epsilon)$
spacetime deviation, multiplied by the large quantity $L$. The motivation
for introducing this kind of variables stems from the detailed discussion
in Ref.\ \cite{Barack}.}
\begin{eqnarray}\label{eqV80}
\Delta_{r }&\equiv&  f_0 L(r _*-r_{*0}), \nonumber\\
\Delta_{r'}&\equiv&  f_0 L(r'_*-r_{*0}), \nonumber\\
\Delta_{t }&\equiv&  f_0 L(t-t_0), \nonumber\\
\Delta_{t'}&\equiv&  f_0 L(t_0-t'), \nonumber\\
 z   &\equiv&  f_0 L \left[(u-u')(v-v')\right]^{1/2},
\end{eqnarray}
where $(r,t)$ is an off-worldline point in the neighborhood of $r_0$,
$(r',t')$ is a worldline point in the past neighborhood of $r_0$,
and $f_0\equiv f^{1/2}(r_0)/r_0$.
We then consider the $l$-mode Green's function as being dependent only on
$L$ and the above ``neutral'' variables, and formally expand the quantities
$\tilde G^{(i)l}_{\alpha'\beta'}$ in powers of $1/L$, while
holding the ``neutral'' variables fixed:
\begin{equation}\label{eqV70}
\tilde G^{(i)l}_{\alpha'\beta'}= \sum_{k=0}^{\infty}L^{-k}
\tilde G^{(i)}_{\alpha'\beta'[k]}(\Delta_r,\Delta_{r'},\Delta_t,z).
\end{equation}
To derive all necessary regularization parameters (including
$D_{\alpha}$), it is sufficient to obtain an expression for the three
leading-order terms in the $1/L$ expansion of the integrand in
Eq.\ (\ref{eqV95}). Higher order terms in this expansion
do not affect the values of the regularization parameters \cite{Barack}.
By analyzing the Green's function equations (\ref{eqV45}), we now show that
the contribution to $\delta F^{(\epsilon)l}_r$ comes at relevant order
only from the three components $\bar G^l_{ttt't'}$, $\bar G^l_{trt't'}$,
and $\bar G^l_{trt'r'}$. In particular, it is shown that the terms of
Eq.\ (\ref{eqV95}) involving $\bar G^l_{rr\alpha'\beta'}$ contribute (for any
$\alpha'\beta'$) only at irrelevant high order.

We start by substituting expansion (\ref{eqV70}) into the
set of Green's function equations (\ref{eqV45}), and pointing out
a few useful ``rules of thumb''\footnote{
In the following discussion we use a terminology according to which
the ``order'' of a mode $\tilde G^{(i)l}_{\alpha'\beta'}$ is determined
by its expansion through Eq.\ (\ref{eqV70}), where the ``neutral''
variables are held fixed. That is, the ``order'' of
$\tilde G^{(i)l}_{\alpha'\beta'}$ is $L^{-k_0}$, where $k_0$ is the
smallest index $k$ for which $\tilde G^{(i)}_{\alpha'\beta'[k]}$ is
non-vanishing.}:
(i) When the operator $D^l_{\rm s}$ acts on a mode
$\tilde G^{(i)l}_{\alpha'\beta'}$ it ``lowers'' its order by 2;
namely, if
$\tilde G^{(i)l}_{\alpha'\beta'}\propto O(L^n)$ (for some $n$), then
$D^l_{\rm s}\tilde G^{(i)l}_{\alpha'\beta'}\propto O(L^{n+2})$.
(ii) A $t$-derivative always acts to lower the order of
$\tilde G^{(i)l}_{\alpha'\beta'}$ by one: if
$\tilde G^{(i)l}_{\alpha'\beta'}\propto O(L^n)$, then
$\tilde G^{(i)l}_{\alpha'\beta',t}\propto O(L^{n+1})$.
(iii) An $r$-derivative may lower the order by one at most
when acting on a function of both $z$ and $\Delta_r$, but does not
affect the order when acting on a function of $z$ alone.\footnote{
This behavior (iii) is special to a turning point. To see this,
note that at $r=r_0$ we have $dz/dr=(f_0/f)L\Delta_{r'}/z$.
Expanding $\Delta_{r'}$ and $z$ in powers of $\tau$ about $\tau(r_0)=0$
and defining the ``neutral'' proper time $\bar\tau$ as in Eq.\
(\ref{eqV210}) below, we find, at $\dot r=0$,
$\Delta_{r'}\cong \frac{1}{2}f_0^{-1}\ddot{r}\bar\tau^2/L$
and $z\cong \bar\tau$ (to leading order in $1/L$).
Consequently, if $\hat f(z)$ is some function of a certain order in
$1/L$, then the $r$-derivative
$d\hat f(z)/dr=[d\hat f(z)/dz](dz/dr)=\frac{1}{2}f^{-1}[d\hat f(z)/dz]
\ddot{r}\bar\tau$ remains of the same order.}
(iv) The source terms $\propto \delta(u-u')\delta(v-v')$ appearing
in Eqs.\ (\ref{eqV45}) are of order $L^2$.
(v) An immediate consequence of all above rules is now apparent from
examining Eqs.\ (\ref{eqV45}): For any given $i$, a $k=0$
contribution to a mode $\tilde G^{(i)l}_{\alpha'\beta'}$ can only come
from the source term (when non-vanishing), whereas the coupling terms
$\tilde {\cal I}^{(i)l}_{(j)}\tilde G^{(j)l}_{\alpha'\beta'}$
contribute only at higher ($k>0$) order.

Consider first Eqs.\ (\ref{eqV45}) for $\alpha'\beta'=t't'$.
In this case, the only sourced equation is the one for
$\tilde G^{(1)l}_{t't'}$. After substituting expansion (\ref{eqV70}),
we find by the above rule (v) that at the $k=0$ order this equation takes
the simple form
$D^l_{\rm s}\tilde G^{(1)l}_{t't'[0]}=\delta(u-u')\delta(v-v')$,
with no contribution from the coupling terms
${\cal I}^{(1)l}_{(j)}\tilde G^{(j)l}_{t't'}$ at this order.
The solution to this equation is of order $L^0$ [see Eq.\ (\ref{eqV200a})
below], thus $\tilde G^{(1)l}_{t't'}\propto O(L^0)$.
Now, the mode $\tilde G^{(2)l}_{t't'}$ is coupled to
$\tilde G^{(1)l}_{t't'}$ through the $\bar G^{(1)}_{,t}$ term in
Eq.\ (\ref{eqIII100(2)}). By the above rules (i) and (ii) we thus find
$\tilde G^{(2)l}_{t't'}\propto O(L^{-1})$.
On the other hand, the mode $\tilde G^{(3)l}_{t't'}$ is excited
only at $k=2$, through coupling with $\tilde G^{(1)l}_{t't'}$
and $\tilde G^{(2)l}_{t't',t}$---see Eq.\ (\ref{eqIII100(3)}).
We find further that
$\tilde G^{(4,5)l}_{t't'}\propto O(L^{-3})$,
$\tilde G^{(6)l}_{t't'}\propto O(L^{-2})$, and
$\tilde G^{(7)l}_{t't'}\propto O(L^{-4})$.
It can be easily checked now that, up to $O(L^{-2})$, the first two
equations of the set (\ref{eqIII95})---the ones with $i=1$ and
$i=2$---form a closed set of equations, with coupling to other
modes affecting only at higher orders:
\begin{equation}\label{eqV100}
\left\{\begin{array}{l}
D_s \tilde G^{(1)l}_{t't'}+\frac{1}{2}ff'\tilde G^{(1)l}_{t't',r}
-\frac{1}{8}\left(f'^2+4ff'/r\right)\tilde G^{(1)l}_{t't'}
+\frac{1}{4}ff' \tilde G^{(2)l}_{t't',t}=\delta(v-v')\delta(u-u'),
\\
D_s \tilde G^{(2)l}_{t't'}+\frac{1}{2}f^{-1}f'\tilde G^{(1)l}_{t't',t}=0.
\end{array}\right.
\end{equation}
Using Eq.\ (\ref{eqV60}) we may now evaluate the order of the $l$-modes
$\bar G^l_{\alpha\beta t't'}$, and proceed to evaluate the order of the
five integrand terms involved in constructing $\delta F_r^{(\epsilon)l}$
through Eq.\ (\ref{eqV95}).
Using rules (ii) and (iii) and recalling $u^{t'}\propto O(L^0)$, we find
that the first integrand term in Eq.\ (\ref{eqV95}) is of order $L^2$,
the second and third terms are of order $L^{1}$, and the last two
terms (the ones involving $\bar G^l_{rrt't'}$) are only of order $L^{-1}$.
Since only integrand terms up to $O(L^0)$ are necessary for calculating
the regularization parameters, we conclude that the component
$\bar G^l_{rrt't'}$ would be of no relevance for this calculation.
Up to the necessary order, the set of Green's function equations
(\ref{eqV45}) therefore reduces, in the case $\alpha'\beta'=t't'$, to
a closed-form set for $\tilde G^{(1)l}_{t't'}$ and $\tilde G^{(2)l}_{t't'}$,
given by Eq.\ ({\ref{eqV100}).

Consider next the set of equations (\ref{eqV45}) for $\alpha'\beta'=t'r'$.
Here, only the $i=2$ component is sourced, and we find
$\tilde G^{(2)l}_{t'r'}\propto O(L^0)$. The modes $i=1$ and $i=3$
are sourced by $\tilde G^{(2)l}_{t'r',t}$, leading to
$\tilde G^{(1,3)l}_{t'r'}\propto O(L^{-1})$.
One similarly finds that the mode $i=4$ is of order $L^{-1}$,
the mode $i=5$ is of order $L^{-2}$, and the modes $i=6,7$
are only of order $L^{-3}$. Again, we use Eq.\ (\ref{eqV60}) to evaluate
the order of each of the five integrand terms appearing in Eq.\
(\ref{eqV95}), this time for $\alpha'\beta'=t'r'$. We now recall,
however, that $u^{r'}$ vanishes at $r_0$, and is therefore of order
$u^{r'}\propto \ddot{r}/L$. It is then easily shown that the only
relevant integrand term is $\bar G^l_{trr't',t}u^{t'}u^{r'}\propto O(L^{ 1})$,
while each of the four other terms contribute to the integral in Eq.\
(\ref{eqV95}) only at order $L^{-2}$, which is irrelevant for
calculating the regularization parameters.
Moreover, one finds that at the relevant order, the mode
$\tilde G^{(2)l}_{t'r'}$ satisfies a single closed-form equation:
\begin{equation}\label{eqV105}
D_s \tilde G^{(2)l}_{t'r'}=\delta(v-v')\delta(u-u').
\end{equation}
(This equation happens to coincide with the scalar Green's function
equation.)

Finally, let us consider Eqs.\ (\ref{eqV45}) for $\alpha'\beta'=r'r'$.
The only sourced mode is now $\tilde G^{(3)l}_{r'r'}$, which is therefore
of order $L^{0}$. However, recalling that at a turning point we have
$(u^{r'})^2\propto O(L^{-2})$, we find that the contribution from this
mode to the integral in Eq.\ (\ref{eqV95}) is only at order $L^{-2}$,
and that the contribution from $\tilde G^{(1)l}_{r'r'}$ and
$\tilde G^{(2)l}_{r'r'}$
is at still higher order. Therefore, no relevant contributions to
$\delta F_r^{(\epsilon)l}$ arise for $\alpha'\beta'=r'r'$.

In conclusion, when restricting our analysis to the case of a turning
point, the problem of calculating the self force via our mode-sum scheme
simplifies considerably:
Instead of the three sets of seven equations each for the $21$ components
required for an arbitrary point of a radial geodesic, one now has to deal
with only the three equations (\ref{eqV100}) and (\ref{eqV105}) (of which
two are coupled and one is closed) for the three components
$\tilde G^{(1)l}_{t't'}$,
$\tilde G^{(2)l}_{t't'}$, and $\tilde G^{(2)l}_{t'r'}$.

Eqs.\ (\ref{eqV100}) and (\ref{eqV105}) are solvable in a
perturbative manner, using the technique described in detail in
Sec.\ V of Ref.\ \cite{Barack}. To apply this technique, one first writes
\begin{equation}\label{eqV190}
\tilde G_{\alpha'\beta'}^{(i)l}=\hat G_{\alpha'\beta'}^{(i)l}
\Theta(u-u')\Theta(v-v') \quad\quad (i=1,2,3),
\end{equation}
where $\Theta$ is the standard step function, acting to confine the
support of the Green's function to within the future light cone of
the source point $x'$. Substituting this expression into Eqs.\ (\ref{eqV100})
and (\ref{eqV105}), one then finds that the new quantities
$\hat G_{\alpha'\beta'}^{(i)l}$ (treated as functions of $v,u$ with a
fixed source point $v',u'$) must satisfy the homogeneous part
of these equations anywhere at $v>v'$ and $u>u'$. One also finds that the
value of $\hat G_{\alpha'\beta'}^{(i)l}$ along the null rays $v=v'$ and
$u=u'$ is uniquely determined from Eqs.\ (\ref{eqV100}) and (\ref{eqV105}).
[Obtaining these ``initial values'' for the quantities
$\hat G_{\alpha'\beta'}^{(i)l}$ involves the solution of a set of ordinary
differential equations along $v=v'$ and $u=u'$ \cite{BL}. In the scalar
case, the Green's function was found, in this way, to admit a constant value
(of unity) along these initial rays \cite{Barack}. In the gravitational
case, the ``initial data'' are a bit more complicated, and will be given
elsewhere \cite{BL}.]
Thus, in effect, the above procedure converts the original
set of Green's function equations into a characteristic initial-data
problem for the quantities $\hat G_{\alpha'\beta'}^{(i)l}$, with a unique
solution. This unique solutions reads, to the relevant order,
\begin{mathletters} \label{eqV200}
\begin{eqnarray}\label{eqV200a}
\hat G_{t't'}^{(1)l}&=&J_0(z)+
\left[f_2(\Delta_r-\Delta_{r'})J_0(z)
      -f_1 (\Delta_r+\Delta_{r'})zJ_1(z) \right] L^{-1} \nonumber\\
&&+\left\{
\left[(\Delta_r-\Delta_{r'})[(f_2^2+f_4)\Delta_r+f_4\Delta_{r'}]
 +f_2^2\Delta_t^2/4\right] J_0(z)
+\left[f_1^2(\Delta_r+\Delta_{r'})^2-f_3\right]z^2 J_2(z)/2
+f_1^2 z^3 J_3(z)/6  \right.\nonumber\\
&&+\left.
\left[-f_1f_2(\Delta_{r}^2-\Delta_{r'}^2)
-f_3(\Delta_{r'}^2+\Delta_r\Delta_{r'}+\Delta_{r'}^2)+f_5 \right]z J_1(z)
\right\}L^{-2}+O(L^{-3}),
\end{eqnarray}
\begin{equation}\label{eqV200b}
\hat G_{t't'}^{(2)l}=-f^{-1}f_2 (\Delta_t+\Delta_{t'}) J_0(z)L^{-1}
+\left[f_{6} z J_1(z)+f_7 J_0(z)\right]\Delta_{r'}\Delta_t L^{-2}
+O(L^{-3}),
\end{equation}
\begin{equation}\label{eqV200c}
\hat G_{t'r'}^{(2)l}=J_0(z) - f_1 z J_1(z)(\Delta_r+\Delta_{r'})
L^{-1} +O(L^{-2}),
\end{equation}
\end{mathletters}
where $J_n(z)$ are the Bessel functions of the first kind,
of order $n$, and the $f_n$'s are radial factors given by
$f_1=\frac{1}{4}f^{-1/2}(rf'-2f)$,
$f_2=rf'f^{-1/2}$,
$f_3=r^2[f^{-1}(f')^2+f'']/12+(f-rf')/2$,
$f_4=r^2f''/2$,
$f_5=f_4+f_2^2/2+rf'/2+1/8$,
$f_6=f^{-1}f_2(2f^{1/2}-f_2)/8$, and
$f_7=r^2[f^{-1}f''-2f^{-2}(f')^2]/4$
(all evaluated at $r=r_0$).

The analysis now proceeds as follows:
(i) We substitute the solutions (\ref{eqV200}) in Eq.\ (\ref{eqV60})
to obtain the relevant $l$-modes $G^{l}_{\alpha\beta\alpha'\beta'}$,
and then substitute these $l$-modes into Eq.\ (\ref{eqV95}).
(ii) To be able to carry out the integration in Eq.\ (\ref{eqV95}),
we next expand all $x'$-dependent quantities now appearing in the
integrand in powers of $\tau$ about the evaluation point $\tau(r_0)=0$.
(This procedure is described in detail, as applied to the analogous
scalar case, in Sec.\ VI of Ref.\ \cite{Barack}.)
(iii) We introduce the ``neutral'' proper time variable\footnote{
Note the different notation used in Ref.\ \cite{Barack}, where the
``neutral'' proper time variable $\bar\tau$ has been denoted by $\lambda$.}
\begin{equation}\label{eqV210}
\bar\tau=-(L/r_0)\tau,
\end{equation}
and write the integrand as an expansion in powers of $1/L$, with
$\bar\tau$ held fixed.

Following these manipulations, Eq.\ (\ref{eqV95}) takes the form
\begin{equation}\label{eqV220}
\delta F_{r}^{(\epsilon)l\pm}=m^2 \int_{0}^{L\epsilon/r_0}
\left[LH_r^{(0)\pm}(\bar\tau)+H_r^{(1)}(\bar\tau)
+H_r^{(2)}(\bar\tau)/L+ O(L^{-2})\right] d\bar\tau,
\end{equation}
where $H_r^{(i)}$ are certain functions of only $\bar\tau$ and $r_0$
(but do not depend on $l$ otherwise). These functions all have the form
of a sum over a few terms $\propto \bar\tau^k J_n(\bar\tau)$, where
$k,n\in\mathbb{N}$ [see, in the analogous scalar case, Eqs.\ (88)--(93)
of Ref.\ \cite{Barack}].
The function $H_r^{(0)}$ has two different values, denoted in Eq.\
(\ref{eqV220}) by $H_r^{(0)+}$ and $H_r^{(0)-}$, according to whether
the derivatives involved in constructing $H_r^{(0)}$ are taken
from $r\to r_0^+$ or $r\to r_0^-$, respectively. This kind of
discontinuity, which shows up only at
the leading order in the $1/L$ expansion, results from differentiating the
$\Theta(u-u')\Theta(v-v')$ factor appearing in Eq.\ (\ref{eqV190}):
the contribution coming from the light cone [through the $\delta(u-u')$
or $\delta(v-v')$ factors] depends, in its overall sign, upon the direction
through which this derivative is taken. This effect is further discussed
and illustrated in Sec.\ IVc of Ref.\ \cite{Barack}
(see especially Fig.\ 1 therein).

In terms of the functions $H_r^{(i)}$, the
regularization parameters are constructed by \cite{Barack}
\begin{equation}\label{eqV230}
A_{r}^{\pm}=m^2 \int_{0}^{\infty}H_r^{(0)\pm}(\bar\tau)d\bar\tau,
\quad\quad
B_{r}=m^2 \int_{0}^{\infty}H_r^{(1)}(\bar\tau)d\bar\tau,
\quad\quad
C_{r}=m^2 \int_{0}^{\infty}H_r^{(2)}(\bar\tau)d\bar\tau,
\end{equation}
and
\begin{eqnarray}\label{eqV240}
D_{r}=-m^2 \lim_{\epsilon\to 0}\sum_{l=0}^{\infty}
\int_{L\epsilon/r_0}^{\infty} \left[LH_r^{(0)\pm}+
H_r^{(1)}+H_r^{(2)}/L \right] d\bar\tau.
\end{eqnarray}
[Both functions $H_r^{(0)+}$ and $H_r^{(0)-}$ can be shown \cite{Barack}
to yield the same contribution to the integral in Eq.\ (\ref{eqV240}),
which is why no $\pm$ sign has been assigned to the parameter $D_{r}$.]
The evaluation of the integrals over $\bar\tau$ [and of the sum over $l$
in Eq.\ (\ref{eqV240})] is done in a manner completely analogous to the
scalar case, as described in Sec.\ VII of \cite{Barack}. Here we merely
give the results of this calculation: The parameters $A^{\pm}_{r}$,
$B_{r}$, $C_{r}$, and $D_{r}$ are found in our
case---a mass particle at a turning point of a radial geodesic---to be
given by
\begin{equation}\label{parameters}
A_r^{\pm}=\mp \frac{m^2}{r_0^2}\left(1-\frac{2M}{r_0}\right)^{-1/2},
\quad\quad
B_r=-\frac{m^2}{2r_0^2},
\quad\quad
C_r=D_r=0.
\end{equation}
We comment that the vanishing of the parameter $C_{r}$ is necessary
to assure consistency of our entire scheme: otherwise, the parameter
$D_{r}$ would have been indefinite (this point is explained in
Sec.\ VIID of \cite{Barack}).\footnote{
Interestingly, the parameter values (\ref{parameters}) coincide
with the values obtained for the {\em scalar} self force acting on a
particle of scalar charge $q=m$, at a turning point of a radial
geodesic in Schwarzschild spacetime---see Ref.\ \cite{Barack}.}

The values derived here for the regularization parameters
find support from a recent numerical analysis carried out by Lousto
\cite{Lousto-Letter,BL}: Lousto has calculated numerically the
bare modes of the force, $F_{\alpha}^{l(\rm{bare})}$, for a radially
moving particle in Schwarzschild spacetime (as part of the
implementation of a different regularization scheme), and found
that these modes indeed show a large-$l$ behavior of the form
indicated in Eq.\ (\ref{eqIV90}) above. Furthermore, the analytic
expressions derived here for the coefficients $A^{\pm}_{r}$, $B_{r}$,
and $C_{r}$ show a perfect agreement with the numerically-derived
coefficients \cite{BL}. In addition, our result of a vanishing parameter
$D_{\alpha}$ confirms, in the cases studied here, Lousto's result based
on his proposed zeta-function regularization. (Although Lousto's numerical
calculations were carried out in a different gauge---in the RW
gauge rather than in the harmonic gauge---this should not alter the values
of the regularization parameters, as we explain in Ref.\ \cite{gauge}.)
Finally, we mention that the expressions derived here for $A^{\pm}_{r}$
and $B_{r}$ also agree with the analytic values obtained recently using
a different formalism, in which the self force is derived from the
so-called Moncrief waveform $\psi$ \cite{BL}.\footnote{
So far, agreement has been achieved for both one-sided values of
$A_{r}$ and $B_{r}$, as well as for the {\em average} value
of $C_{r}$. Currently, we obtain, using Moncrief's formalism, that
although the averaged $C_{r}$ vanishes (as in our present analysis),
the one-sided values of $C_{r}$ fail to vanish. It is most likely that
this preliminary result is erroneous. This point awaits further examination.}

In conclusion of this section, we have found that the $r$ component of the
regularized self force at a turning point of a radial geodesic
in Schwarzschild spacetime is given by above Eq.\ (\ref{eqV40}), where the
parameter $B_{r}$ is given in Eq.\ (\ref{parameters}), and where
$\bar F^{l{\rm (bare)}}_{r}$ are the (sided average) $l$-modes of the
bare force. These bare modes are derived [through the second equality of Eq.\
(\ref{eqIV5})] from the metric perturbation {\em in the harmonic gauge}.
For practical use, it would be desirable to express our result in a more useful
gauge (e.g., in the RW gauge). This shall be done in a forthcoming
paper \cite{gauge}, as part of a more general discussion of the gauge issue
in the context of the gravitational self force problem.


\section{Summary and discussion}\label{secVI}

In this manuscript we have generalized the scheme of mode-sum
regularization, previously applied only in the scalar case, to the
problem of calculating the gravitational self-force on a mass particle.
The proposed scheme offers a practical way for implementing the formal
prescription developed by MSTQW, even in strong-field calculations.

Within the mode-sum scheme, the basic formula for constructing the
(harmonic-gauge related) gravitational self-force is given by
\begin{equation}\label{eqVI10}
F_{\alpha }^{\rm{H}}=\sum_{l=0}^{\infty }\left( [F_{\alpha }^{l\rm{
(bare)}}]^{\rm{H}}-A_{\alpha }L-B_{\alpha }-C_{\alpha }/L\right)
-D_{\alpha },
\end{equation}
where $L\equiv l+1/2$ and the label `H' indicates a quantity associated
with the metric perturbation in the harmonic gauge.
To apply this formula in actual calculations (i.e., for a certain orbit on
a specific background), one needs to be provided with
(i) the bare force modes $[F_{\alpha}^{l(\rm bare)}]^{\rm H}$ and
(ii) the values of the four regularization parameters $A_{\alpha}$,
$B_{\alpha}$, $C_{\alpha}$, and $D_{\alpha}$.
To obtain the bare modes $[F_{\alpha}^{l(\rm bare)}]^{\rm H}$, one
first calculates the multipole modes of the metric
perturbation in the harmonic gauge, and then uses the relation
\begin{equation}\label{eqVI20}
[F_{\alpha }^{l\rm{(bare)}}]^{\rm{H}}=
m k_{\alpha}{}^{\beta\gamma\delta}[\bar h_{\beta\gamma}^l]_{;\delta}^{\rm H},
\end{equation}
where $k^{\alpha\beta\gamma\delta}$ is the tensor given in Eq.\
(\ref{eqII50}) and $[\bar h_{\beta\gamma}^l]^{\rm H}$ is the $l$ mode of
the trace-reversed metric perturbation (in the harmonic gauge).
Whereas this part of the analysis---the derivation of the bare
modes---typically involves a numerical analysis, the derivation of
the regularization parameters may be carried out analytically,
by means of local analysis of the Green's function. This kind of local analysis
was described and demonstrated in Sec.\ \ref{secV}, where we
constructed all necessary regularization parameters for two simple cases.

The first case examined in Sec.\ \ref{secV} was the trivial test case of
a static particle in flat spacetime. Here, the mode-sum scheme easily
reproduced the obvious result of a vanishing self force.
We then calculated the regularization parameters for the case of
a mass particle at a turning point of a radial geodesic in Schwarzschild
spacetime. The values of these parameters were given in Eq.\ (\ref{parameters}).
These values find support from a recent numerical analysis by Lousto
\cite{Lousto-Letter,BL}.

Our calculation scheme---like the formal prescription by MSTQW on which it
relies---is formulated within the harmonic gauge.
In an accompanying paper \cite{gauge} we explore the gauge dependence of
the gravitational self force, and re-formulate our scheme in terms
of other gauges---ones more commonly adopted in perturbation analysis
(such as the RW gauge). We then conclude that an expression of the
form (\ref{eqVI10}) is applicable for calculating the self force
in any specific gauge `G' (as long as the the self force attains a
definite finite value in that gauge), by just replacing the harmonic
gauge modes $[F_{\alpha }^{l\rm{(bare)}}]^{\rm{H}}$ on the right-hand
side of Eq.\ (\ref{eqVI10}) with the `G'-gauge modes
$[F_{\alpha }^{l\rm{(bare)}}]^{\rm{G}}$---the ones derived from the
G-gauge metric perturbation using Eq.\ (\ref{eqVI20}), with `H'$\to$ 'G'.
The analysis of Ref.\ \cite{gauge} further tells us that the regularization
parameters in Eqs.\ (\ref{eqVI10}) should not carry
any gauge label: these parameters are ``gauge invariant'', in the sense
that they are always to be derived from the Green's function
associated with the harmonic-gauge wave operator [the one defined through
Eq.\ (\ref{eqII60})], irrespective of the gauge in which the bare
modes are calculated. In particular, we find that the values specified in
Eq.\ (\ref{parameters}) are valid under any gauge.

\subsection*{Further applications of the proposed calculation scheme}

The analysis of Sec.\ \ref{secIII} provides formal tools for calculating
the gravitational self force in any orbit on a Schwarzschild background.
For any such orbit, the regularization parameters may be derived by local
analysis of the Green's function modes, based on the separated system of
equations (\ref{eqIII95}) [supplemented by Eqs.\ (\ref{eqIII100})]---as
demonstrated in Sec.\ \ref{secV}.
It should be noted, however, that in more general cases than the
simple ones considered here, the derivation of the regularization
parameters shall require much more technical effort: Here, we
only had to deal with one equation for one Green's function component
(in the case of a static particle in flat spacetime), or with three
coupled equations for three components (in the case of a particle at
a turning point of a radial geodesic). We shall have to face $3$ sets
of $7$ coupled equations each for a total of $21$ components already for
an arbitrary point of a radial geodesic.
In general cases, one would have to deal with up to 58 equations
(7 sets of 7 coupled equations each for even perturbation modes,
and 3 sets of 3 equations each for odd perturbation modes).

To fully implement the regularization scheme and derive the self force,
one also needs to calculate the bare modes of the force.
This was already done by Lousto \cite{Lousto-Letter}, using
Moncrief's formalism \cite{Moncrief}, for radial geodesic motion in
Schwarzschild spacetime. In his analysis, Lousto calculated numerically
the (one-sided averaged) values of $A_{\alpha}$, $B_{\alpha}$, and
$C_{\alpha}$, and presumed a null value for $D_{\alpha}$ based on a proposed
zeta-function regularization scheme. To provide a full theoretical basis for
Lousto's results in the radial motion case (and verify its consistency with
MSTQW regularization) we intend to calculate analytically, using our mode-sum
scheme, all four parameters $A_{\alpha}$, $B_{\alpha}$, $C_{\alpha}$, and
$D_{\alpha}$ for an arbitrary point of a radial geodesic \cite{BL}.
Then, the next natural step would be to consider more general, non-radial
orbits. This would require a considerable amount of both analytic work
(deriving the regularization parameters) and numerical work (calculating the
bare modes of the force by solving the metric perturbation equations).

We finally comment on the applicability of our mode-sum regularization
scheme to orbits in Kerr spacetime. Although the theoretical basis for
applying our scheme for non-spherically symmetric backgrounds
has not been developed yet (not even in the scalar case), we believe that a
properly generalized version of this scheme could, eventually, cope with the
Kerr case as well. Such a generalization could still be based on MSTQW's formal
prescription (\ref{eqII40}), which applies for any vacuum spacetime.
The main obstacle in re-designing our scheme for the Kerr case
would be, of course, the non-separability of the metric perturbation and
Green's function into multipole modes in the time domain (such a separation
has been a necessary step when executing our scheme in spherically
symmetric cases). This difficulty would make both the analytic and
numerical parts of the mode-sum scheme more challenging:
The bare modes of the force would usually have to be
calculated in the frequency domain (using the Teukolsky/Chrzanowski formalism
\cite{Chrzanowski}), and then appropriately summed over Fourier frequencies.
As to the analytic part of the scheme, it seems to us that, with the use
of an appropriate local perturbation analysis,
enough information for constructing the regularization parameters could be
extracted from the time-domain Green's function equations, even without fully
separating these equations.

\section*{Acknowledgments}

I am very grateful to Amos Ori for his guidance throughout the
execution of this research project, and for many interesting discussions.
It is a pleasure to thank Eric Poisson for his valuable comments on an
early version of this manuscript.
This research was partially supported by the United States--Israel
Binational Science Foundation;
and by a Marie Curie Fellowship of the European Community programme
IHP-MCIF-99-1 under contract number HPMF-CT-2000-00851.

\appendix

\section{Coupling of Green's function's components
(Schwarzschild spacetime)}

In this appendix we give explicitly the coupling terms appearing
in the Green's function equation (\ref{eqII70}), for the
Schwarzschild spacetime case (and using Schwarzschild coordinates).
In the following expressions $f\equiv 1-2M/r$ and a prime denotes $d/dr$.

\begin{equation} \label{App1a}
{\cal A}_{tt}^{\mu\nu}\bar G_{\mu\nu}=
\left[f^{-1}(f')^2/2-2f'\partial_r\right]   \bar G_{tt}
+2f'\partial_t                              \bar G_{tr}
+\left[f^2f''-f(f')^2/2\right]              \bar G_{rr}
+r^{-3}ff' \left(                           \bar G_{\theta\theta}+
\sin^{-2}\theta                             \bar G_{\varphi\varphi}
\right),
\end{equation}
\begin{equation} \label{App1b}
{\cal A}_{tr}^{\mu\nu}\bar G_{\mu\nu}=
\left[f''-f^{-1}(f')^2-2r^{-2}f\right]      \bar G_{tr}
+f^{-2}f'\partial_t                         \bar G_{tt}
+f'\partial_t                               \bar G_{rr}
-2r^{-3}(\partial_{\theta}+\cot\theta)       \bar G_{t\theta}
-2r^{-3}\sin^{-2}\theta\,\partial_\varphi    \bar G_{t\varphi},
\end{equation}
\begin{eqnarray} \label{App1c}
{\cal A}_{rr}^{\mu\nu}\bar G_{\mu\nu}&=&
\left[2f'\partial_r-4f/r^2+f^{-1}(f')^2/2\right]    \bar G_{rr}
+2f^{-2}f'\partial_t                                \bar G_{tr}
+\left[f^{-2}f''-f^{-3}(f')^2/2\right]              \bar G_{tt}\nonumber\\
&&-4r^{-3}(\partial_{\theta}+\cot\theta)            \bar G_{r\theta}
-4r^{-3}\sin^{-2}\theta\,\partial_\varphi           \bar G_{r\varphi}
+\left(2r^{-4}-r^{-3}f^{-1}f'\right)\left(          \bar G_{\theta\theta}
+\sin^{-2}\theta                                    \bar G_{\varphi\varphi}
\right),
\end{eqnarray}
\begin{eqnarray} \label{App1d}
{\cal A}_{t\theta}^{\mu\nu}\bar G_{\mu\nu}=
\left[f'/r-(f'+2f/r)\partial_r-r^{-2}\sin^{-2}\theta\right] \bar G_{t\theta}
+f'\partial_t                                               \bar G_{r\theta}
+2(f/r)\partial_{\theta}                                    \bar G_{tr}
-2r^{-2}\sin^{-2}\theta\cot\theta\,\partial_\varphi         \bar G_{t\varphi},
\end{eqnarray}
\begin{eqnarray} \label{App1e}
{\cal A}_{r\theta}^{\mu\nu}\bar G_{\mu\nu}&=&
\left[(f'-2f/r)\partial_r-4f/r^2-r^{-2}\sin^{-2}\theta\right] \bar G_{r\theta}
+2(f/r)\partial_{\theta}                                      \bar G_{rr}
+f^{-2}f'\partial_t                                           \bar G_{t\theta}
-2r^{-3}(\partial_{\theta}+\cot\theta)                        \bar G_{\theta\theta}
\nonumber\\&&
+2r^{-3}\sin^{-2}\theta\cot\theta                             \bar G_{\varphi\varphi}
-2r^{-2}\sin^{-2}\theta\cot\theta\,\partial_{\varphi}         \bar G_{r\varphi}
-2r^{-3}\sin^{-2}\theta\,\partial_{\varphi}                   \bar
G_{\theta\varphi},
\end{eqnarray}
\begin{eqnarray} \label{App1f}
{\cal A}_{\theta\theta}^{\mu\nu}\bar G_{\mu\nu}&=&
\left[-4(f/r)\partial_r+2(f/r^2)-2r^{-2}\sin^{-2}\theta\right]  \bar G_{\theta\theta}
+rf^{-1}f'                                                      \bar G_{tt}
+(2f^2-rff')                                                    \bar G_{rr}
+4(f/r)\partial_{\theta}                                        \bar G_{r\theta}
\nonumber\\&&
+2(r^{-2}\cot^2\theta+f'/r)\sin^{-2}\theta                      \bar G_{\varphi\varphi}
-4r^{-2}\sin^{-2}\theta\cot\theta\,\partial_{\varphi}           \bar
G_{\theta\varphi},
\end{eqnarray}
\begin{eqnarray} \label{App1g}
{\cal A}_{\varphi\varphi}^{\mu\nu}\bar G_{\mu\nu}&=&
\left[-4(f/r)\partial_r-4r^{-2}\cot\theta\,\partial_{\theta}+2(f/r^2)
    +2r^{-2}\cot^2\theta\right]                         \bar G_{\varphi\varphi}
+rf^{-1}f'\sin^{2}\theta                                \bar G_{tt}
+(2f^2-rff')\sin^{2}\theta                              \bar G_{rr}
\nonumber\\&&
+4(f/r)\sin\theta \cos\theta                            \bar G_{r\theta}
+4(f/r)\partial_{\varphi}                               \bar G_{r\varphi}
+\left[2r^{-2}\cos^2\theta+2(f'/r)\sin^{2}\theta\right] \bar G_{\theta\theta}
+4r^{-2}\cot\theta\,\partial_\varphi                    \bar G_{\theta\varphi},
\end{eqnarray}
\begin{eqnarray} \label{App1h}
{\cal A}_{t\varphi}^{\mu\nu}\bar G_{\mu\nu}=
\left[-(f'+2f/r)\partial_r-2r^{-2}\cot\theta\partial_{\theta}+2f'/r\right]
                                                    \bar G_{t\varphi}
+2r^{-2}\cot\theta\partial_{\varphi}                \bar G_{t\theta}
+2(f/r)\partial_{\varphi}                           \bar G_{tr}
+f'\partial_t                                       \bar G_{r\varphi},
\end{eqnarray}
\begin{eqnarray} \label{App1i}
{\cal A}_{r\varphi}^{\mu\nu}\bar G_{\mu\nu}&=&
\left[(f'-2f/r)\partial_r-4f/r^2-2r^{-2}\cot\theta\,\partial_{\theta}\right]
                                                \bar G_{r\varphi}
+2(f/r)\partial_{\varphi}                       \bar G_{rr}
+f^{-2}f'\partial_t                             \bar G_{t\varphi}
-2r^{-3}\sin^{-2}\theta\,\partial_{\varphi}     \bar G_{\varphi\varphi}
\nonumber\\&&
+2r^{-2}\cot\theta\,\partial_{\varphi}          \bar G_{r\theta}
-2r^{-3}(\partial_\theta + \cot\theta)          \bar G_{\theta\varphi},
\end{eqnarray}
\begin{eqnarray} \label{App1j}
{\cal A}_{\theta\varphi}^{\mu\nu}\bar G_{\mu\nu}&=&
\left[-4(f/r)\partial_r-2r^{-2}\cot\theta\,\partial_{\theta}
-3r^{-2}\cot^2\theta-2f'/r+(2f-1)/r^2\right]            \bar G_{\theta\varphi}
+2(f/r)\partial_{\varphi}                               \bar G_{r\theta}
\nonumber\\&&
+2(f/r)(\partial_\theta-2\cot\theta)                    \bar G_{r\varphi}
+2r^{-2}\cot\theta\,\partial_{\varphi}                  \bar G_{\theta\theta}
-2r^{-2}\sin^{-2}\theta\,\cot\theta\,\partial_{\varphi} \bar
G_{\varphi\varphi}.
\end{eqnarray}




\begin{references}
\bibitem{QWerror} Note the mistake in Eqs.\ (49)--(51) of Ref.\ \cite{QW},
where the trace-reversed Green's function has been used instead of the Green's
function associated with the metric perturbation itself.
\end{references}
\end{document}